\documentclass[a4paper,11pt]{article}
\pdfoutput=1 % if your are submitting a pdflatex (i.e. if you have
\usepackage{jheppub} % for details on the use of the package, please
\usepackage[T1]{fontenc} % if needed

\title{\boldmath Anomalous decay rate of quasinormal modes in Schwarzschild-dS and Schwarzschild-AdS black holes}

\author[a]{Almendra Arag\'{o}n}
\author[a]{P. A. Gonz\'{a}lez}
\author[b]{Eleftherios Papantonopoulos}
\author[c]{Yerko V\'asquez}

\affiliation[a]{Facultad de
Ingenier\'{i}a y Ciencias, Universidad Diego Portales, \\ Avenida Ej\'{e}rcito
Libertador 441, Casilla 298-V, Santiago, Chile.}
\affiliation[b]{Physics Division,
National Technical University of Athens, \\15780 Zografou Campus,
Athens, Greece.}
\affiliation[c]{Departamento de F\'isica y Astronom\'ia, Facultad de Ciencias, Universidad de La Serena,\\
Avenida Cisternas 1200, La Serena, Chile.}

\emailAdd{almendra.aragon@mail.udp.cl}
\emailAdd{pablo.gonzalez@udp.cl}
\emailAdd{lpapa@central.ntua.gr}
\emailAdd{yvasquez@userena.cl}

\abstract{Recently  an anomalous decay rate of the quasinormal modes of a massive scalar field in Schwarzschild black holes backgrounds was reported in  which the longest-lived modes are the ones with higher angular number,
for a scalar field mass smaller than
a critical value, while that beyond this value the 
behaviour is inverted.
In this work, we extend the study to other asymptotic geometries, such as, Schwarzschild-de Sitter and Schwarzschild-AdS black holes. Mainly, we found that such behaviour and the critical mass are present in  the Schwarzschild-de Sitter background. 
Also, we found that the value of the critical mass increases when the cosmological constant increases and also when the overtone number is increasing. On the other hand, despite the critical mass
is not present in Schwarzschild-AdS black holes backgrounds, the decay rate of the quasinormal modes always exhibits an anomalous behaviour.}

\begin{document} 
\maketitle
\flushbottom

\section{Introduction}

The quasinormal modes (QNMs) and quasinormal frequencies (QNFs)
\cite{Regge:1957td,Zerilli:1971wd, Kokkotas:1999bd, Nollert:1999ji, Konoplya:2011qq,Berti:2009kk}
have recently acquired great interest due
to the detection of gravitational waves \cite{Abbott:2016blz}.
Despite the detected signal is consistent with the Einstein gravity \cite{TheLIGOScientific:2016src}, there are possibilities for alternative theories of gravity due to the large uncertainties in mass and angular momenta of the ringing black hole \cite{Konoplya:2016pmh}. The QNMs and QNFs give
information about the stability of matter fields that
evolve perturbatively in the exterior region of a black hole without backreacting on the
metric. Also, the QNMs are characterized by a spectrum that is independent of the initial conditions of the perturbation and depends on the black hole parameters and probe field parameters, and on the fundamental constants of the system. The QNM infinite discrete spectrum consists of complex frequencies, $\omega=\omega_R+i \omega_I$, in which the real part $\omega_R$ determines the oscillation timescale of the modes, while the complex part $\omega_I$ determines their exponential decaying timescale (for a review on QNM modes see \cite{Kokkotas:1999bd, Berti:2009kk}).

The QNFs have been calculated by means of numerical and analytical techniques; some well known numerical methods are: the Mashhoon method, Chandrasekhar-Detweiler method, WKB method, Frobenius method, method of continued fractions, Nollert, asymptotic iteration method (AIM) and improved AIM, among others. In the case of gravitational perturbations it was found that for the Schwarzschild and Kerr black hole backgrounds the longest-lived modes are always the ones with lower angular number $\ell$.  This is expected in a physical system because the more energetic modes with high angular number $\ell$ would have faster decaying rates. In the case of a massive probe scalar field it was found \cite{Konoplya:2004wg, Konoplya:2006br,Dolan:2007mj, Tattersall:2018nve},  at least for the overtone $n = 0$, that if we have a light scalar field, then the longest-lived quasinormal modes are those with a high angular number $\ell$, whereas for a heavy scalar field the longest-lived modes are those with a low angular number $\ell$.
This behaviour can be understood because for the case of massive scalar field even if its mass is small its fluctuations can maintain the quasinormal modes to live longer even if the angular number $\ell$ is large.  This anomalous behaviour is depending on whether the mass of the scalar field exceeds a critical value or not. This anomalous decay rate for small mass scale of the scalar field was recently discussed in \cite{Lagos:2020oek}.

Extensive study of QNMs of black holes in
asymptotically flat spacetimes have been performed for the last few decade mainly due to the potential astrophysical interest. Considering the case when the black hole is
immersed in an expanding universe, the QNMs of black holes in de Sitter (dS) space have been investigated \cite{deSitter_1,deSitter_2}.
The  AdS/CFT correspondence \cite{Maldacena:1997re,Aharony:1999ti} stimulated the interest in calculating the QNMs and QNFs of  black holes in  anti-de Sitter (AdS) spacetimes. It was shown in \cite{Horowitz:1999jd} that  this principle leads to a   correspondence of the QNMs of the  gravity bulk to the decay of perturbations in the dual conformal field theory, and the QNFs were also studied in Ref. \cite{Chan:1996yk}.

The aim of this work is to study the propagation of scalar fields  in the Schwarzschild-dS  and Schwarzschild-AdS black hole backgrounds in order to see  if there is an anomalous decay rate of quasinormal modes. The motivation for this study is to study the effect of the presence of the cosmological constant on the anomalous decay. For gravitational perturbations in Schwarzschild de Sitter black holes the longest-lived modes are always the one with lower angular number $\ell$ \cite{Otsuki}. In the case of a massive probe scalar field, as we already discussed, a different behaviour in the Schwarzschild background was observed, here a mass scale is introduced by the probe scalar field. However,  the cosmological constant is introducing another scale and it is interesting to see what is the effect on the anomalous behaviour of the competition of the two scales for both positive and negative cosmological constant. We carry out this study by using the pseudospectral Chebyshev method \cite{Boyd}
which is an effective method to find high overtone modes  \cite{Finazzo:2016psx,Gonzalez:2017shu,Gonzalez:2018xrq,Becar:2019hwk,Aragon:2020qdc}.

The gravitational QNMs of Schwarzschild-de Sitter black hole were studied in \cite{ Mellor:1989ac, Otsuki, Moss:2001ga}. The QNMs  of a probe scalar field for this geometry were calculated in \cite{Zhidenko:2003wq} by using the sixth order WKB formula and the approximation by the P\"oschl-Teller (P-T) potential. Also, it was shown the frequencies all have a negative imaginary part, which means that the propagation of scalar fields is stable in this background. The presence of the cosmological constant leads to decrease of the real oscillation frequency and to a slower decay, and  high overtones was studied in Ref. \cite{Konoplya:2004uk}. Also, a novel infinite set of purely imaginary modes was found \cite{Jansen:2017oag}, which depending on the black hole mass may even be the dominant mode.

In the case of a massless scalar field in the background of a Schwarzschild-dS  black hole we find two types of QNMs, a family of complex quasinormal modes  which are well described by the WKB formula (photon sphere modes) and  a family of pure imaginary ones, which are closely related to the de Sitter horizon (dS modes) \cite{Cardoso:2017soq}. These modes have a  different behaviour  as the cosmological constant is changing. First of all for the complex modes  all the frequencies  have a negative imaginary part, which means that the propagation of scalar field is stable in this background. However the presence of a larger cosmological constant leads to decrease the real oscillation frequency and to a slower decay. On the contrary in the case of pure imaginary modes we find the cosmological constant leads to a fast decay, when it increases, that is, contrary to the complex QNFs.

In the case of a massive scalar field in the background of a Schwarzschild-dS  black hole we find that the imaginary part of the photon sphere QNFs has an anomalous behaviour for a scalar field mass less than a critical mass, i.e, the absolute value of the imaginary part decay  when the angular harmonic numbers $\ell$ increase; and for a scalar field mass greater than the critical mass the behavior is inverted, i.e,  the longest-lived modes are always the ones with higher angular number. The critical mass corresponds to the value of the scalar field mass where the behavior of the decay rate of the QNMs is inverted and can be obtained from the condition $Im(\omega)_{\ell}=Im(\omega)_{\ell+1}$ in the {\it eikonal} limit, that is when $\ell \rightarrow \infty$.
Additionally, we find that as the value of
the cosmological
constant increases the value of the critical mass also increases. 

On the other hand, we do not observe an anomalous decay for the dS modes. We also show that the dS modes can acquire a real part which depends on the scalar fields mass. The QNMs of massive scalar fields in the Schwarzschild-dS black hole background were studied in  \cite{Toshmatov:2017qrq} showing that there 
a lower limit of the scalar mass which allows the waves with QNFs to propagate at infinity. In the case of Schwarzschild-AdS black hole background we find 
a faster decay when the mass of the scalar field increases and when the angular harmonic numbers $\ell$ decrease.

The manuscript is organized as follows: In Sec. \ref{QNM}, we study the scalar field stability by calculating  the QNFs of scalar perturbations numerically of a massless and massive  scalar field in the background of Schwarzschild-dS and Schwarzschild-AdS black hole background by using the pseudospectral Chebyshev method. In Sec. \ref{WKBJ}  we perform an analysis using the WKB method to get some analytical insight. We conclude in Sec. \ref{conclusion}.

\section{Scalar perturbations}
\label{QNM}

The Schwarzschild-(dS)AdS black holes
 are maximally symmetric solutions of the equations of motion that arise from the action
\begin{equation}
    S=\frac{1}{16\pi G}\int d^4x\sqrt{-g}(R-2\Lambda)\,,
\end{equation}
where $G$ is the Newton constant, $R$ is the Ricci scalar and $\Lambda$ the cosmological constant. The Schwarzschild-dS and Schwarzschild-AdS black holes are described by the metric
\begin{equation}
    ds^2=f(r)dt^2-\frac{dr^2}{f(r)}-r^2(d\theta^2+sin^2\theta d\phi^2)\,,
    \label{metric}
\end{equation}
where $f(r)=1-\frac{2M}{r}-\frac{\Lambda r^2}{3}$, $M$ is the black hole mass, $\Lambda > 0$ in the metric represents  the Schwarzschild-dS black hole, while $\Lambda < 0$ represents  the Schwarzschild-AdS black hole. For the Schwarzschild-dS black hole the difference between the cosmological horizon $r_{\Lambda}$ and the event horizon $r_H$ decreases when the cosmological constant increases, and both horizons coincide when $9 M
^2 \Lambda  =1$, while for the  Schwarzschild-AdS black hole there is only one horizon that decreases when the absolute value of the cosmological constant increases.

The QNMs of scalar perturbations in the background of the metric (\ref{metric})
are given by the scalar field solution of the Klein-Gordon equation
\begin{equation}
\frac{1}{\sqrt{-g}}\partial _{\mu }\left( \sqrt{-g}g^{\mu \nu }\partial_{\nu } \varphi \right) =-m^{2}\varphi \,,  \label{KGNM}
\end{equation}%
with suitable boundary conditions for a black hole geometry. In the above expression $m$ is the mass
of the scalar field $\varphi $. Now, by means of the following ansatz
\begin{equation}
\varphi =e^{-i\omega t} R(r) Y(\Omega) \,,\label{wave}
\end{equation}%
the Klein-Gordon equation reduces to
\begin{equation}
\frac{1}{r^2}\frac{d}{dr}\left(r^2 f(r)\frac{dR}{dr}\right)+\left(\frac{\omega^2}{f(r)}+\frac{\kappa}{r^2}-m^{2}\right) R(r)=0\,, \label{radial}
\end{equation}%
where we defined $\kappa=-\ell (\ell+1)$, with $\ell=0,1,2,...$, which represents the eigenvalue of the Laplacian on the two-sphere and $\ell$ is the multipole number.
Now, defining $R(r)=\frac{F(r)}{r}$
and by using the tortoise coordinate $r^*$ given by
$dr^*=\frac{dr}{f(r)}$,
 the Klein-Gordon equation can be written as a one-dimensional Schr\"{o}dinger-like equation
 \begin{equation}\label{ggg}
 \frac{d^{2}F(r^*)}{dr^{*2}}-V_{eff}(r)F(r^*)=-\omega^{2}F(r^*)\,,
 \end{equation}
 with an effective potential $V_{eff}(r)$, which  parametrically thought,  $V_{eff}(r^*)$, is given  by
  \begin{equation}\label{pot}
 V_{eff}(r)=-\frac{f(r)}{r^2} \left(\kappa -  m^2 r^2-f^\prime(r)r\right)~.
 \end{equation}

\subsection{Numerical analysis. Schwarzchild-de Sitter black holes. }

Now, in order to compute the QNFs, we will solve numerically the differential equation (\ref{radial}) by using the pseudospectral Chebyshev method, see for instance \cite{Boyd}. First, it is convenient to perform a change of variable in order to limit the values of the radial coordinate to the range $[0,1]$. Thus, we define the change of variable $y=(r-r_H)/(r_{\Lambda}-r_H)$. So, the event horizon is located at $y=0$ and the cosmological horizon at $y=1$.
The radial equation (\ref{radial}) becomes
\begin{eqnarray} \label{rad}
&&\nonumber f(y) R''(y) + \left( \frac{2 \left(    r_{\Lambda}- r_H  \right) f(y)}{r_H+\left( r_{\Lambda}-r_H \right) y } + f'(y) \right) R'(y)\\
&&+ \left( r_{\Lambda}-r_H  \right)^2 \left( \frac{\omega^2}{f(y)}- \frac{ \ell(\ell+1)}{\left( r_H + \left( r_{\Lambda}-r_H \right)y \right)^2} -m^2  \right) R(y)=0\,.
\end{eqnarray}
In the vicinity of the horizon (y $\rightarrow$ 0) the function $R(y)$ behaves as
\begin{equation}
R(y)=C_1 e^{-\frac{i \omega \left( r_{\Lambda}-r_H \right)}{f'(0)} \ln{y}}+C_2 e^{\frac{i \omega \left( r_{\Lambda}-r_H \right)}{f'(0)} \ln{y}} \,.
\end{equation}
Here, the first term represents an ingoing wave and the second represents an outgoing wave near the black hole horizon.
So, imposing the requirement of only ingoing waves at the horizon, we fix $C_2=0$. On the other hand, at the cosmological horizon the function $R(y)$ behaves as
\begin{equation}
R(y)= D_1 e^{-\frac{i \omega \left( r_{\Lambda}-r_H \right)}{f'(1)} \ln{(1-y)}}+D_2 e^{\frac{i \omega \left( r_{\Lambda}-r_H \right)}{f'(1)} \ln{(1-y)}}  \,.
\end{equation}
Here, the first term represents an outgoing wave and the second represents an ingoing wave near the cosmological horizon. So, imposing the requirement of only ingoing waves on the cosmological horizon requires $D_1=0$.
Taking the behaviour of the scalar field at the event and cosmological horizons we define the following ansatz
\begin{equation}
R(y)= e^{-\frac{i \omega \left( r_{\Lambda}-r_H \right)}{f'(0)} \ln{y}} e^{\frac{i \omega  \left( r_{\Lambda}-r_H \right)  }{f'(1)} \ln{(1-y)}} F(y) \,.
\end{equation}
Then, by inserting the above ansatz for $R(y)$ in Eq. (\ref{rad}), an differential equation for the function $F(y)$ is obtained. The solution for the function $F(y)$ is assumed to be a finite linear combination of the Chebyshev polynomials, and it is inserted into the differential equation for $F(y)$. Also, the interval $[0,1]$ is discretized at the Chebyshev collocation points. Then, the differential equation is evaluated at each collocation point. So, a system of algebraic equations is obtained, and it corresponds to a generalized eigenvalue problem, which is solved numerically to obtain the QNFs ($\omega$). In appendix \ref{Accuracy} we show the accuracy of the numerical technique used.

\begin{figure}[h!]
\begin{center}
\includegraphics[width=0.44\textwidth]{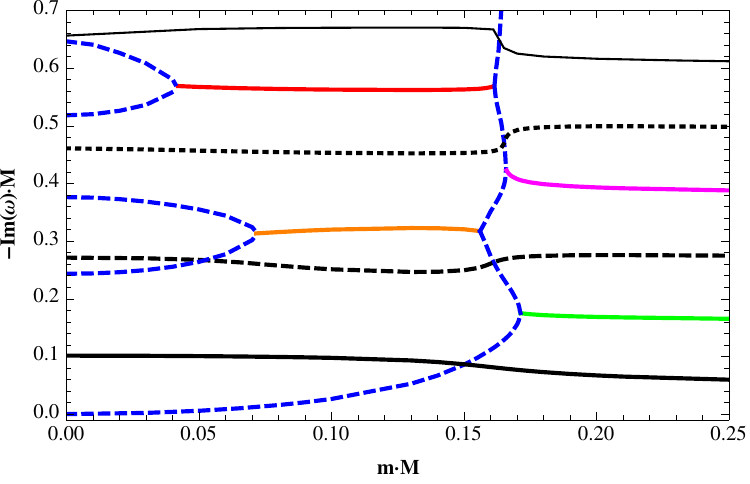}
\includegraphics[width=0.45\textwidth]{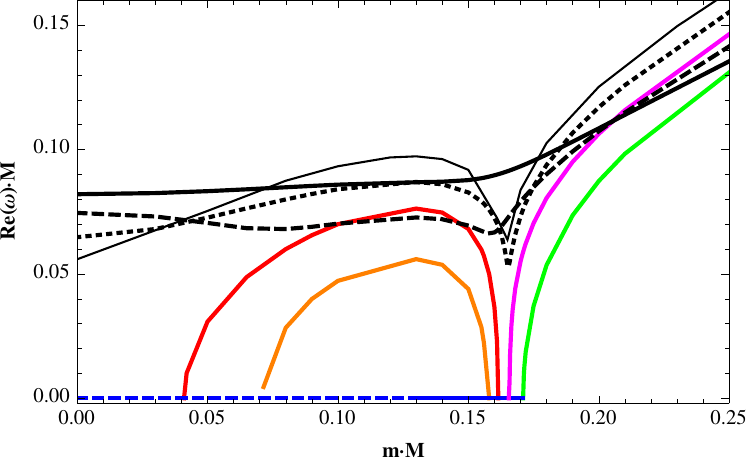}
\end{center}
\caption{The behaviour of the QNFs, 
%the imaginary part of the quasinormal frequencies
$-Im(\omega)M$ (left panel), and 
%the real part of the quasinormal frequencies 
$Re(\omega)M$ (right panel), as a function of 
%the scalar field mass 
$mM$ for different overtone numbers, with $\ell=0$, %$M=1$, 
and $M^2\Lambda=0.04$.}
\label{F9}
\end{figure}

Now, in order to visualize the different families of QNMs we plot in Fig. \ref{F9} the behaviour of
 $-Im(\omega)M$ (left panel), and $Re(\omega)M$
(right panel) as a function of $mM$
for different overtone numbers and $\ell=0$.
In these figures we can recognize two families, for zero mass, a family of complex QNFs given by the black curves, and a purely imaginary family given by the blue dashed curves. The purely imaginary modes belong to the family of de Sitter modes, in that they continuously
approach those of empty de Sitter in the limit that the black hole vanishes, 
while the complex ones are those of the Schwarzschild black hole, in the sense
that they limit to the modes of the asymptotically 
at Schwarzschild black
hole in that limit, and this family corresponds to the photon sphere modes.
Also, we observe the behavior of the families when $mM$
the scalar field mass 
increases. We see that the photon sphere modes (black curves), is the dominant family for $mM > 0.15$.
Interestingly, the dS modes (colored curves) is the family dominant for $mM < 0.15$; however, for some value of $mM$,
the purely imaginary de Sitter modes (blue dashed curves) can acquire a real part, given by the continuous colored curves. In the next subsections, we study the QNFs for the photon sphere modes and for the dS modes separately.

\clearpage

\subsubsection{Photon sphere modes}

\begin{figure}[h]
\begin{center}
\includegraphics[width=0.42\textwidth]{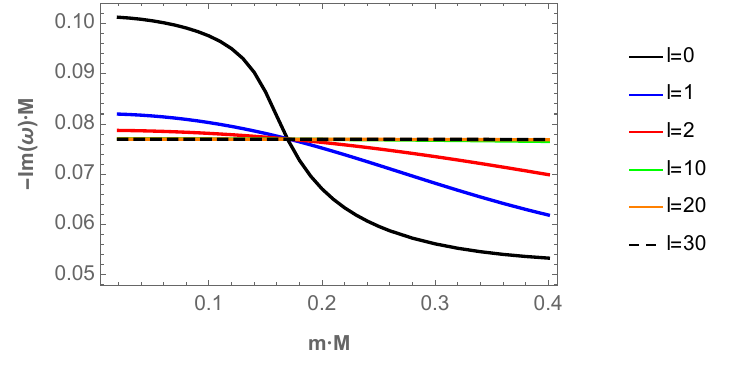}
\includegraphics[width=0.42\textwidth]{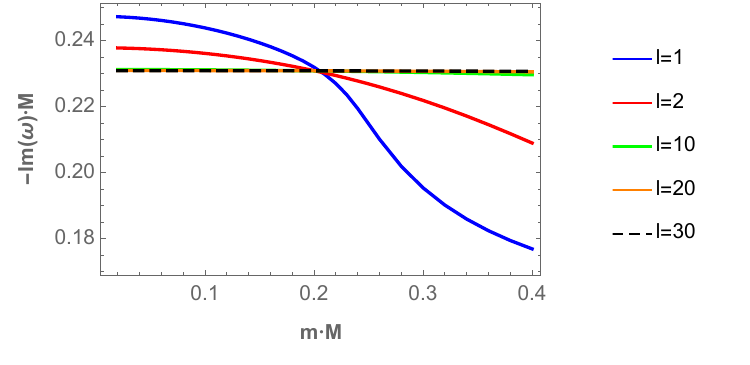}
\includegraphics[width=0.38\textwidth]{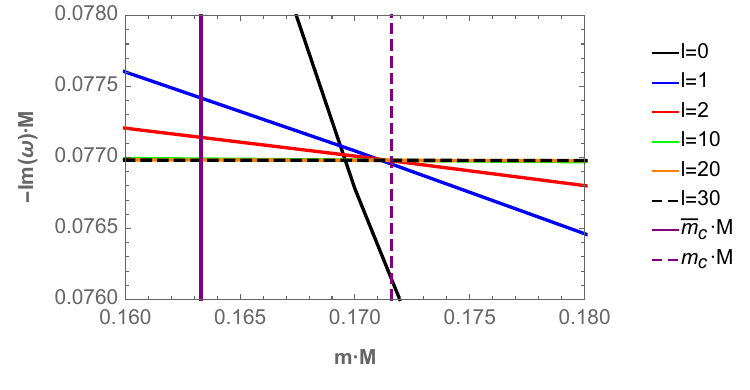}
\end{center}
\caption{Top panel for the behaviour of 
$Im(\omega) M$ of the photon sphere modes, 
top-left panel for the dominant mode $n_{PS}=0$, and right top-right panel for the first overtone $n_{PS}=1$,
as a function of $mM$ for different values of the parameter $\ell=0,1,2,10,20,30$ (top-left panel), and $\ell=1,2,10,20,30 \ge n_{PS}$ (top-right panel), with 
$M^2\Lambda=0.04$. Bottom panel corresponds to a zoom of the top-left panel, for the range 
$0.16<mM<0.18$, $\bar{m}_cM\approx 0.1633 $, and $m_cM\approx 0.1716$ (WKB).}
\label{F1}
\end{figure}

Now, in order to show the existence of anomalous decay rate of QNMs, we plot in Fig. \ref{F1} the behaviour of the fundamental QNFs  of this family (top-left panel), for different values of the multipole number $\ell$, and different values of $mM$.
The numerical values are in appendix \ref{tables} Table \ref{T1}. It is possible to observe that the imaginary part of these frequencies has an anomalous behaviour for $mM <m_cM$, i.e, the absolute value of the imaginary part decay when the angular harmonic numbers $\ell$ increase; however, for $mM>m_cM$ the behaviour is normal, i.e, the longest-lived modes are the ones with smaller angular number. 
 The critical mass $m_c$ corresponds to 
the value of the scalar field mass where the behaviour in the QNFs is inverted, and can be obtained from the condition $Im(\omega)_{\ell}=Im(\omega)_{\ell+1}$ in the {\it eikonal} limit, that is when $\ell \rightarrow \infty$. Looking at Fig. \ref{F9} combined with Fig. \ref{F1}  shows that the anomalous decay rate occurs at subdominant order for $mM<0.15$; however for $0.15<mM<m_cM$ the anomalous decay rate occurs at dominant order. Also, in order to show the same anomalous behaviour for the first overtone number $n_{PS}=1$, we plot in Fig. \ref{F1}, the behaviour of $-Im(\omega)M$ of the photon sphere modes (top-right panel), for $\ell \geq n_{PS}$ as a function of $mM$. \footnote{We have left outside the case for $\ell=0$, because the imaginary part of the QNFs exhibits a behaviour other than $\ell>0$, that is, there is a region, where the absolute value of $-Im(\omega)M$
increases when $mM$ increases, see Fig. \ref{F9}, and the curve for $\ell=0$ does not intersect the other curves, for the first overtone number ($n_{PS}=1$). While that, for the dominant mode ($n_{PS}=0$) the behaviour is opposite, and always is the same, that is, the absolute value of $-Im(\omega)M$
always decreases, see Fig. \ref{F9}, and the curve for $\ell=0$ intersects the other curves. In general, this different behaviour occurs for $\ell < n_{PS}$.} Note that, the critical mass value increases when the overtone number $n_{PS}$ increases, for $\ell \geq n_{PS}$. The numerical values are in appendix \ref{tables} Table \ref{T3}. Also, the behaviour of the real
part of the QNFs is smooth, and there is a slower decay of the mode when $mM$ 
increases, see appendix \ref{tables} Table \ref{T1} and \ref{T3}. Also, it is observed that all the QNFs have negative imaginary part, which means that the propagation of scalar field is stable in this background. The bottom panel of Fig. \ref{F1} corresponds to a zoom of the top-left panel, we can appreciate better that for high values of $\ell$ the QNFs cross very close at the critical mass.

\begin{figure}[h]
\begin{center}
\includegraphics[width=0.5\textwidth]{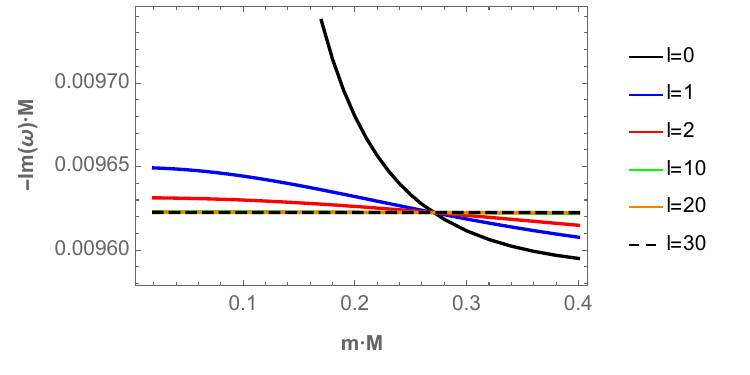}\\
\includegraphics[width=0.5\textwidth]{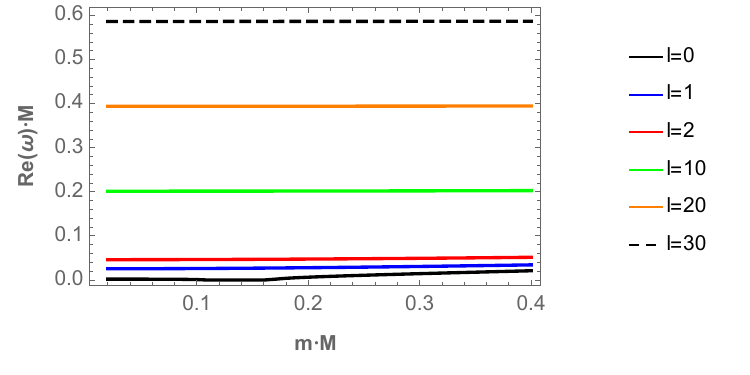}
\end{center}
\caption{The behaviour of the QNFs 
%$Im(\omega) M$,
for the dominant photon sphere modes  ($n_{PS}=0$),
$Im(\omega) M$ (top panel)  and $Re(\omega)M$ (bottom panel), as a function of  $mM$ for different values of the parameters $\ell=0,1,2,10,20,30$, with 
%$M=1$, and 
$M^2\Lambda=0.11$.}
\label{F100}
\end{figure}

Now, in order to see the influence of the cosmological constant in the critical mass, we plot in Fig. \ref{F100} the behaviour of the fundamental QNFs, for different values of the multipole number $\ell$, and different values of the mass of the scalar field, but for a cosmological constant greater than the previous case $M^2\Lambda=0.11$. The numerical values are in appendix \ref{Tables2} Table \ref{T6}. We can observe that for a greater cosmological constant the value of the critical mass increases, this will be confirmed analytically in the next section with the WKB formula.

It is interesting to note that, despite that the spacetime is asymptotically dS, where the boundary conditions are imposed at the event horizon and at the cosmological horizon the effective potential tends to $-\Lambda(3m^2-2\Lambda)r^2/9$  for $\ell=0$, at infinity, and it can diverge positively, negatively, or be null, specifically, it vanishes for $m=\bar{m}_c=\pm \sqrt{2\Lambda/3}$. So, for $M
^2\Lambda=0.04$, and for a scalar field with mass $mM=\bar{m}_cM\approx 0.163$, and also for $M^2\Lambda=0.11$ and for a scalar field with mass $mM=\bar{m}_cM\approx 0.278$, the effective potential vanishes at infinity. 
%These are the critical masses we have considered  with $n=0$. 
So, for a scalar field with $m=\bar{m}_c$,
%critical mass, 
and $\ell=0$ the effective potential at infinity is not divergent.  Also, for $\ell\neq 0$ and $m=\bar{m}_c$, the effective potential tends to a negative constant at infinity given by $-\ell (\ell+1) \Lambda /3$ and the scalar field does not generate such divergence. It is worth mention this 
%the critical 
scalar field mass ($m=\bar{m}_c$) also satisfies the lower limit on the value of the mass of the scalar field for the waves with QNFs related to the effective potential barrier between the black hole and the cosmological horizons of the Schwarzschild dS spacetime, and enabling the wave to reach observer at infinity without meeting any other potential wall proposed in Ref. \cite{Toshmatov:2017qrq}.
The physical picture behind it is that there is a specific critical scale of the scalar field that cancels out the scale introduced by the cosmological constant.

\subsubsection{dS modes}
\begin{figure}[h]
\begin{center}
\includegraphics[width=0.44\textwidth]{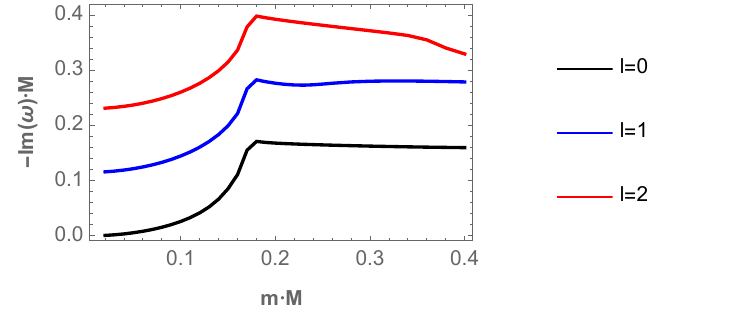}
\includegraphics[width=0.44\textwidth]{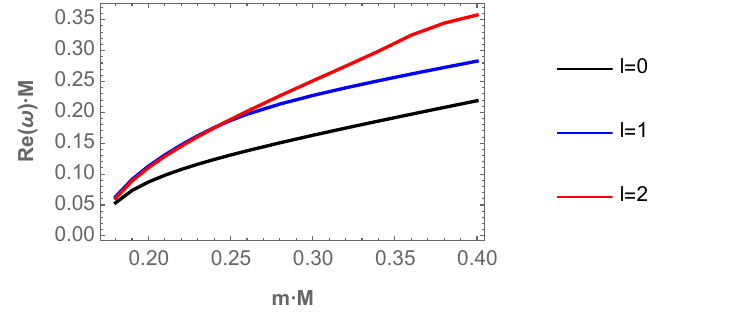}
\end{center}
\caption{The behaviour of the dominant dS modes ($n_{dS}=0$), 
$-Im(\omega)M$ (left pannel) and $Re(\omega)M$ (right pannel) as a function of
%the scalar field mass 
$mM$ for different values of the parameter $\ell=0,1,2$, with 
%$M=1$, and 
$M^2\Lambda=0.04$.}
\label{F2}
\end{figure}

These modes continuously
approach those of empty de Sitter spacetime in the limit that the black hole vanishes, and the QNFs are given by \cite{LopezOrtega:2006my}
\begin{equation}
\omega = -i \sqrt{\frac{\Lambda}{3}} \left(  \frac{3}{2} + \ell +2n_{dS} \pm \left(  \frac{9}{4} - \frac{3 m^2}{\Lambda}\right)^{1/2}  \right) \,,
\end{equation}
for $m^2< 3 \Lambda /4$ and
\begin{equation}
\omega = \pm \sqrt{\frac{\Lambda}{3}}\left(  -\frac{9}{4} + \frac{3 m^2}{\Lambda}\right)^{1/2}-i \sqrt{\frac{\Lambda}{3}} \left(  \frac{3}{2} + \ell +2n_{dS}   \right) \,,
\end{equation}
for $m^2> 3 \Lambda /4$, where $n_{dS}=0,1,2, \dots$. Note that, for massless scalar field the QNFs are purely imaginary; however, for $m^2> 3 \Lambda /4$, the QNFs acquire a real part. The behaviour of the fundamental dS modes is showed in Fig. \ref{F2} as a function of $mM$ for different values of the multipole number and $M^2 \Lambda=0.04$. We can observe that the modes of this family can acquire a real part depending on the scalar field mass, see Table \ref{T2}.  This happens approximately for $mM \approx 0.17$ in Fig. \ref{F2}. Thus, as it was observed all the QNFs are negative, which means that the propagation of scalar field is stable in this background.
Moreover, we can observe a faster decay when the parameter $\ell$ increases, and a faster decay when the scalar field mass increases until  the QNFs acquire a real part after which the decay is stabilized. Also, the real part increases when the scalar field mass increases. Thus, for the range of mass considered, the longest-lived modes are always the ones with lowest angular number, and the anomalous behaviour is not observed for this family of modes.

%\clearpage

%\begin{figure}[h!]
%\begin{center}
%\includegraphics[width=0.5\textwidth]{Re.pdf}
%\end{center}
%\caption{The behaviour of the real part of the quasinormal frequencies $Re(\omega)$ as a function of the scalar field mass $m$ for different overtone numbers, with $\ell=0$, $M=1$, and $\Lambda=0.04$.}
%\label{Real11}
%\end{figure}

%\clearpage

\subsection{Numerical analysis. Schwarzchild-AdS black holes.}

In this case,
%It is  convenient to compare our result with those of \cite{Horowitz:1999jd}, so we express the mass $M$ as a function of the event horizon $r_H$
%\begin{equation}
%M=-\frac{\Lambda r_H^3}{6}+ \frac{r_H}{2}\,,
%\end{equation}
%where the cosmological constant is taken as $\Lambda=-\frac{3}{R^2}$, with $R$ being the AdS radius. Now,
under the change of variable $y=1-r_H/r$ the radial equation (\ref{radial}) becomes
\begin{equation} \label{r}
(1-y)^4 f(y) R''(y) +(1-y)^4 f'(y)R'(y) + \left( \frac{\omega^2 r_H^2}{f(y)}- \ell(\ell+1) (1-y)^2 -m^2 r_H^2 \right) R(y)=0\, ,
\end{equation}
where the prime denotes derivative with respect to $y$. In the new coordinate the event horizon is located at $y=0$ and the spatial infinity at $y=1$. In the neighborhood of the horizon (y $\rightarrow$ 0) the function $R(y)$ behaves as
\begin{equation}
R(y)=C_1 e^{-\frac{i \omega r_H}{f'(0)} \ln{y}}+C_2 e^{\frac{i \omega r_H}{f'(0)} \ln{y}} \,,
\end{equation}
where the first term represents an ingoing wave and the second represents an outgoing wave near the black hole horizon. Imposing the requirement of only ingoing waves on the horizon, we fix $C_2=0$. Also, at infinity the function $R(y)$ behaves as
\begin{equation}
R(y)= D_1 (1-y)^{\frac{3}{2} + \sqrt{\left(\frac{3}{2}  \right)^2 -\frac{ 3 m^2}{\Lambda}}}+ D_2 (1-y)^{\frac{3}{2} - \sqrt{\left(\frac{3}{2}  \right)^2 -\frac{3 m^2}{\Lambda}}} \,.
\end{equation}
So, imposing the scalar field vanishes at infinity requires $D_2=0$. Therefore, by considering the behaviour at the event horizon and at infinity of the scalar field, it is possible to define the following ansatz
\begin{equation}
R(y)= e^{-\frac{i \omega r_H}{f'(0)} \ln{y}} (1-y)^{\frac{3}{2} + \sqrt{\left(\frac{3}{2}  \right)^2 -\frac{3 m^2}{\Lambda}}} F(y) \,.
\end{equation}

\begin{figure}[h]
\begin{center}
\includegraphics[width=0.45\textwidth]{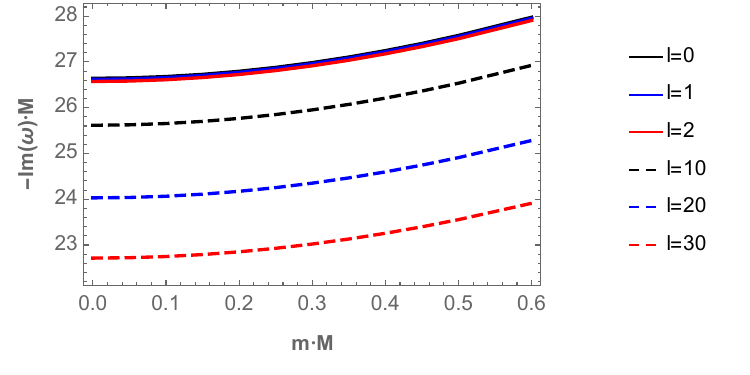}
\includegraphics[width=0.45\textwidth]{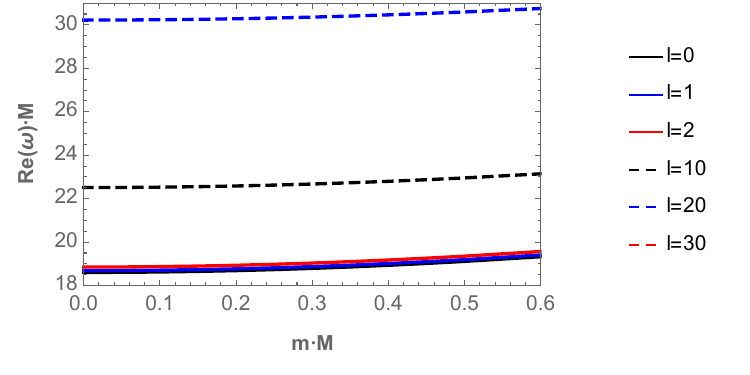}
\end{center}
\caption{The behaviour of fundamental mode $-Im(\omega)M$ (left figure)  and $Re(\omega)M$ (right figure) as a function of $mM$ for different values of the parameter $\ell=0,1,2,10,20,30$, with $M^2\Lambda=-765075$.}
\label{F1B}
\end{figure}

Then, by inserting this last expression in Eq. (\ref{r}) we obtain an equation for the function $F(y)$, which we solve numerically employing the pseudospectral Chebyshev method, as in the previous case. 
Also, in order to see if there is an anomalous decay rate of quasinormal modes, we plot in Fig. \ref{F1B} the behaviour of the  fundamental QNFs, for different values of the parameter $\ell$, and different values of $mM$.
The numerical values are in Appendix \ref{C} Table \ref{T7}. We can observe an anomalous behaviour of the QNMs in Schwarzschild-AdS black holes because the longest-lived modes are always the ones with higher angular number; however, there is not a critical mass where the behaviour of the modes is inverted. The same behaviour is observed for 
%small values of 
{\bf{$M^2\Lambda=-1$}}, see  Appendix \ref{C} Table \ref{T72}.
Also, they present a faster decay when the scalar field mass increases and when the angular harmonic numbers $\ell$ decrease. The frequency of the oscillations increases sightly  when the scalar field mass increases  and also when the angular harmonic numbers $\ell$ decrease. On the other hand, the same behaviour is observed for small values of $M^2\Lambda$, and small $\ell$, see Fig. \ref{F1B1}, the numerical values are in Appendix \ref{C} Table \ref{T721}, where we found some QNFs for low values of $\ell=0,1,2$, however, for high values of $\ell$ the numerical method is not efficient, requiring a very large number of Chebyshev polynomials to converge due to the imaginary part becomes very small, which corresponds to weakly damped modes predicted by \cite{ 
  Grain:2006dg, Festuccia:2008zx}, and analyzed further in \cite{Berti:2009wx}. It is worth mentioning that despite the analysis with respect to the critical mass was performed in the eikonal limit,
even for small values of $\ell$ for the Schwarzschild dS case there is a reversal of the behaviour of the decay rate of the QNFs at a value of the mass very close to the critical mass of the scalar field, Fig. \ref{F1}.
We have not observed an reversal of the behaviour of the decay rate of the QNFs for Schwarzschild AdS black hole for small values of $\ell$, see Fig. \ref{F1B1}, and therefore we would not expect an inverted behaviour for high values of $\ell$ at some critical mass. However, it would be worthwhile to check this claim with some numerical method that works well in this limit, such as the method used in  Ref. \cite{Berti:2009wx}.
\begin{figure}[h]
\begin{center}
\includegraphics[width=0.47\textwidth]{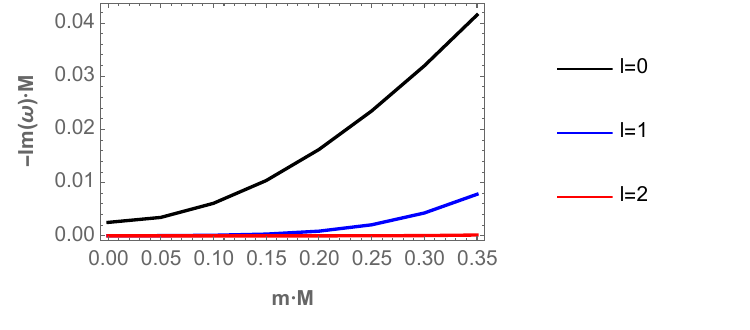}
\includegraphics[width=0.47\textwidth]{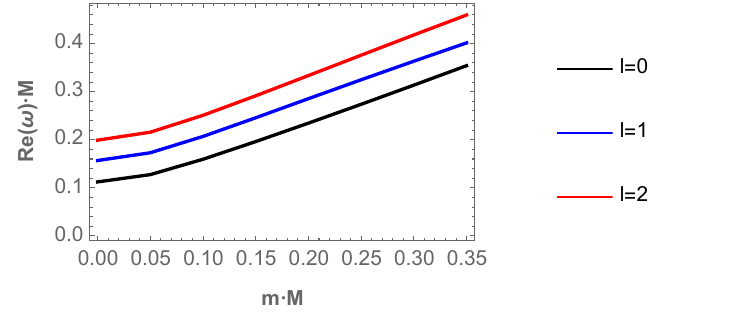}
\end{center}
\caption{The behaviour of fundamental mode $-Im(\omega)M$ (left figure)  and $Re(\omega)M$ (right figure) as a function of $mM$ for different values of the parameter $\ell=0,1,2$, with $M^2\Lambda=-0.005$.}
\label{F1B1}
\end{figure}

In this case, 
the effective potential at infinity always diverges, due to the fact that the cosmological constant is negative, and consequently the scalar field can probe the divergence of the effective potential at infinity.

\section{Analysis using the WKB method}
\label{WKBJ}

In this section, in order to get some analytical insight of the behavior of the QNFs in the eikonal limit $\ell \rightarrow \infty$, we use the method based on Wentzel-Kramers-Brillouin (WKB) approximation initiated by Mashhoon \cite{Mashhoon} and by Schutz and Iyer \cite{Schutz:1985zz}. Iyer and Will computed the third order correction \cite{Iyer:1986np}, and then it was
%Konoplya 
extended to the sixth order \cite{Konoplya:2003ii}, and recently up to the 13th order \cite{Matyjasek:2017psv}, see also \cite{Konoplya:2019hlu}. 
%has been done by Matyjasek and Opala [12] (see also [13]).

This method has been used to determine the QNFs for asymptotically flat and asymptotically de Sitter black holes. This is due to the WKB method can be used for effective potentials which have the form of potential barriers that approach to a constant value at the horizon and spatial infinity \cite{Konoplya:2011qq}, so, it does apply to determine the QNFs of Schwarzschild (de-Sitter) black hole; however
%however, it {\bf generically} does not apply to the Schwazschild AdS black hole.
only the photon sphere modes can be obtained with this method. The QNMs are determined by the behavior of the effective potential near its maximum value $r^*_{max}$. The Taylor series expansion of the potential around its maximum is given by
\begin{equation}
V(r^*)= V(r^*_{max})+ \sum_{i=2}^{i=\infty} \frac{V^{(i)}}{i!} (r^*-r^*_{max})^{i} \,.
\end{equation}
where
\begin{equation}
V^{(i)}= \frac{d^{i}}{d r^{*i}}V(r^*)|_{r^*=r^*_{max}}\,,
\end{equation}
corresponds to the $i$-th derivative of the potential with respect to $r^*$ evaluated at the maximum of the potential $r^*_{max}$. Using the WKB approximation up to 6th order the QNFs are given by the following expression \cite{Hatsuda:2019eoj}

\begin{eqnarray}
\omega^2 &=& V(r^*_{max})  -2 i U \,,
\end{eqnarray}
where
\begin{eqnarray}
\notag U &=&  N\sqrt{-V^{(2)}/2}+\frac{i}{64} \left( -\frac{1}{9}\frac{V^{(3)2}}{V^{(2)2}} (7+60N^2)+\frac{V^{(4)}}{V^{(2)}}(1+4 N^2) \right)  \\
\notag && +\frac{N}{2^{3/2} 288} \Bigg( \frac{5}{24} \frac{V^{(3)4}}{(-V^{(2)})^{9/2}} (77+188N^2) +\frac{3}{4} \frac{V^{(3)2} V^{(4)}}{(-V^{(2)})^{7/2}}(51+100N^2)\\
\notag &&  +\frac{1}{8} \frac{V^{(4)2}}{(-V^{(2)})^{5/2}}(67+68 N^2)+\frac{V^{(3)}V^{(5)}}{(-V^{(2)})^{5/2}}(19+28N^2)+\frac{V^{(6)}}{(-V^{(2)})^{3/2}} (5+4N^2)  \Bigg)\,,
\end{eqnarray}
and $N=n_{PS}+1/2$, with $n_{PS}=0,1,2,\dots$, is the overtone number. The imaginary and real part of the QNFs can be written as
\begin{eqnarray}
\label{im} \omega_I^2 &=& - (Im(U)+V/2)+\sqrt{(Im(U)+V/2)^2+Re(U)^2} \,, \\
\omega_R^2 &=& -Re(U)^2 / \omega_I^2 \,,
\end{eqnarray}
respectively, where $Re(U)$ is the real part of $U$ and $Im(U)$ its imaginary part.

Defining $L^2= \ell (\ell+1)$, we find that for large values of $L$, the maximum of the potential is approximately at
\begin{equation}
r_{max} \approx 3M-\frac{M}{3 L^2} (1-27 m^2 M^2 +18 \Lambda M^2) \mathcal{B} \,,
\end{equation}
and
\begin{equation} \label{coa}
V(r^*_{max}) \approx \mathcal{B}\left( \frac{L^2}{27 M^2}+\frac{2+ 27 m^2 M^2 -18 \Lambda M^2}{81 M^2} \right) \,,
\end{equation}
where
\begin{equation}
\mathcal{B}= 1-9 \Lambda M^2\,,
\end{equation}
while the second derivative of the potential evaluated at $r^*_{max}$ yields
%\begin{equation}
%\resizebox{1\hsize}{!}{
%V^{(2)}(r^*_{max}) \approx \mathcal{B}^2 %\left(-\frac{2 L^2}{729 M^4}+ \frac{2(-8+54m^2 M^2 -18 \Lambda M^2-1215 m^2 M^4 \Lambda+810 \Lambda^2 M^4 )}{6561 M^4} \right)} \,.
%\end{equation}
\begin{equation}
V^{(2)}(r^*_{max}) \approx \mathcal{B}^2 \left(-\frac{2 L^2}{729 M^4}- \frac{2 \mathcal{A}}{6561 M^4} \right)\,,
\end{equation}
where
\begin{equation}
    \mathcal{A}= 8-54 m^2 M^2+18 \Lambda M^2 +1215 m^2 M^4 \Lambda -810 \Lambda^2 M^4\,.
\end{equation}

For the higher derivatives, only the leading terms are important in the limit considered

\begin{eqnarray}
V^{(3)}(r^*_{max})  && \approx \mathcal{B}^3 \frac{4 L^2}{6561 M^5}\,, \\
V^{(4)}(r^*_{max}) && \approx \mathcal{B}^3 \frac{16 L^2}{19683 M^6}\,, \\
V^{(5)}(r^*_{max}) && \approx -\mathcal{B}^4 \frac{40 L^2}{59049 M^7}\,, \\
V^{(6)}(r^*_{max}) && \approx -\mathcal{B}^4 \frac{16 L^2 (4+15 \Lambda M^2)}{177147 M^8}\,. \label{cof}
\end{eqnarray}

Using these results we find that $U$ evaluated at $r^*_{max}$ is given approximately by

%\begin{eqnarray}
%\notag U &\approx & (1-9 \Lambda M^2) \Bigg[ \frac{(1+2n) L}{54 M^2}+i\frac{-65+99 \Lambda M^2 + 6 n (n+1) (-23+45 \Lambda M^2)}{5832 M^2} +\frac{1+2n}{209952 M^2 L} \Bigg( 749 \\
%\notag && -5n (n+1)(1-9 \Lambda M^2)(61+423 \Lambda M^2) -11664 m^2 M^2 +262440 m^2 M^4 \Lambda +6930 \Lambda M^2 -162405 \Lambda^2 M^4 \Bigg) \\
%\notag &&  -i\frac{(-38+99 \Lambda M^2+6n(n+1) (-14+45 \Lambda M^2))(8-54 m^2 M^2+18 \Lambda M^2 +1215 m^2 M^4 \Lambda -810 \Lambda^2 M^4)}{26244 M^2 L^2}\Bigg] \,.
%\end{eqnarray}

\begin{eqnarray}
\notag U &\approx &\mathcal{B} \Bigg[ \frac{(1+2n_{PS}) L}{54 M^2}+i\frac{-65+99 \Lambda M^2 + 6n_{PS}(n_{PS}+1)(-23+ 45 \Lambda M^2)}{5832 M^2} +\frac{1+2n_{PS}}{209952 M^2 L} \Bigg( 749 \\ 
\notag && 
 -5n_{PS} (n_{PS}+1)\mathcal{B}(61+423 \Lambda M^2) -11664 m^2 M^2+ 262440 m^2 M^4 \Lambda +6930 \Lambda M^2 \\ 
 && -162405 \Lambda^2 M^4\Bigg)-i\frac{(-38+99 \Lambda M^2+6n_{PS}(n_{PS}+1) (-14+45 \Lambda M^2))\mathcal{A}}{26244 M^2 L^2}\Bigg]\,,
\end{eqnarray}
thereby, by using Eq. (\ref{im}) we obtain
\begin{eqnarray}
\label{EE}
\notag\omega_I^2 &\approx& \frac{(1+2n_{PS})^2\mathcal{B}}{M^2} \Bigg( \frac{1}{108} \\
+&&\frac{\mathcal{B}(137-29160 m^2 M^2 +18045 \Lambda M^2 +235 n_{PS}(n_{PS}+1)\mathcal{B})}{209952 L^2} \Bigg).
\end{eqnarray}
%where
%\begin{equation}
%\mathcal{B}= 1-9 \Lambda M^2\,.
%\end{equation}
The second term of this expression is zero at the value of the critical mass $m_c$, which is given by
\begin{equation}\label{mc}
m_c M = \sqrt{\frac{137+18045 \Lambda M^2 +235 n_{PS} (n_{PS}+1)(1-9 \Lambda M^2)}{29160}} \,.
\end{equation}
Therefore, we can observe an inverted behavior of $\omega_I(\ell)$. For $m>m_c$, $\omega_I$ increases with $\ell$; whereas, for $m<m_c$, $\omega_I$ decreases when $\ell$ increases.

\begin{figure}[h]
\begin{center}
\includegraphics[width=0.6\textwidth]{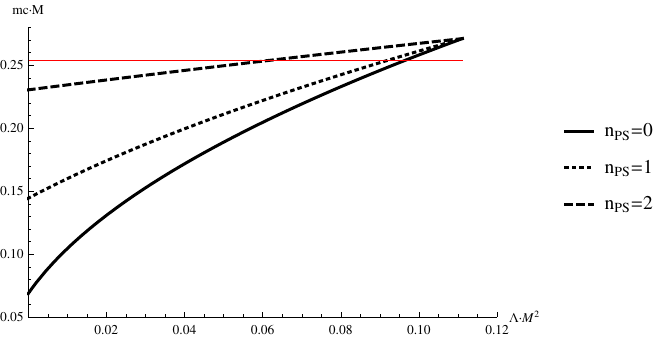}
\end{center}
\caption{The behaviour of $m_c M$ as a function of $\Lambda M^2$ for different values of the overtone number $n_{PS}=0,1,$ and $2$. For values of the critical mass below the horizontal red line the zero mode is subdominant at the critical mass, while it is dominant at the critical mass for values above the line.} 
\label{FMC}
\end{figure}

For $\Lambda=0$ we recover the Eq. (28) of Ref. \cite{Lagos:2020oek}. For $\Lambda M^2 = 0.04$ and $n_{PS}=0$ we obtain from (\ref{mc}) the value $m_c M= 0.1716$ and for $n_{PS}=1$ we obtain $m_c M= 0.1994$, which agrees with the numerical results shown in Fig. (\ref{F1}) for high values of $\ell$. For $\Lambda M^2 =0.11$ and $n_{PS}=0$ we obtain $m_c M=0.2698$ which agrees with the numerical result shown in Fig. (\ref{F100}). In Fig. \ref{FMC}, we plot the behaviour of $m_c M$ as a function of $M^2\Lambda$ for different values of the overtone number. We can see that, for small values of $M^2\Lambda$ the critical mass depends on the overtone number $n_{PS}$, that is, the critical mass value increases when the overtone number increases. However, when $M^2\Lambda$ increases, the black hole becomes extremal when $9 M
^2 \Lambda=1$, and the critical mass value does not depend on the overtone number, the same behaviour was observed in the context of $f(R)$ gravity \cite{Aragon:2020xtm}.  We have found numerically, using the pseudospectral method, that the $\ell=0$ mode that goes to a zero mode in the zero mass limit is subdominant for critical mass $m_c M < m_* M$, while it is dominant for $m_c M > m_* M$, where $m_* M \approx 0.2544$ (which corresponds to $\Lambda M
^2 \approx 0.097$ and $n_{PS}=0$). The horizontal red line of Fig. \ref{FMC} corresponds to $m_c M = m_* M$, therefore, for values of critical mass below the line, the zero mode is subdominant at the critical mass, while it is dominant at the critical mass for values above the line. Note that naively extrapolating the WKB result Eq. (\ref{mc}) for the critical mass to negative $\Lambda$
 \footnote{We thank the referee  for pointing out this fact to us.},
 it is possible to obtain a positive critical mass for $\Lambda M^2 > -0.007$ and $n_{PS}=0$, which
suggests that could exist a range of small AdS curvature with a nonzero critical mass. Thus, for $\Lambda M^2=-0.005$, and $n_{PS}=0$, WKB analysis could predict a critical mass at $m_cM \approx 0.04$. However, as we mentioned in the previous subsection, we did not find any indication of a critical mass for Schwarzschild AdS black hole.

\newpage

Now, we perform a analysis similar to Ref. \cite{Lagos:2020oek} to show an inverted behavior of the width of the potential in the case of Schwarzschild de-Sitter black hole. The coefficients $V^{(i)}$ of the Taylor expansion of the potential around its maximum in the large $\ell$ expansion can be written as
\begin{equation}
V^{(i)}= A^{(i)} L^2 +B^{(i)}\,,
\end{equation}
where the leading coefficients $A^{(i)}$ are given in Eqs. (\ref{coa})-(\ref{cof}). Thus, we obtain
\begin{equation}
%\resizebox{.9\hsize}{!}{
\omega_I^2= \frac{\mathcal{B}}{108 M^2}-\frac{5 \mathcal{B}^3 (59+279 \Lambda M^2)+52488 M^2 \mathcal{B} B^{(0)}+708588 M^4 B^{(2)}}{209952 M^2\mathcal{B} L^2}\,.
%}
\end{equation}
This Eq. can be expressed as
\begin{equation}
\omega_I^2= -\frac{1}{8}\frac{V^{(2)}}{V(r^*_{max})}-\frac{5 \mathcal{B}^2 (59+279 \Lambda M^2)}{209952 M^2 L^2}\,.
\end{equation}
In this expression $V$ and $V^{(2)}$ consider their leading and sub-leading order terms when $\ell \rightarrow \infty$. The first term of this expression can be written in terms of the maximum height of the potential and its width around $r^*_{max}$. This can be seen expanding the potential around $r^*_{max}$ up to second order derivatives:
\begin{equation}
V(r^*) \approx V(r^*_{max}) + \frac{V^{(2)}}{2} (r^*-r^*_{max})^2\,.
\end{equation}
From here, considering an $r^*$ near $r^*_{max}$ such that $V(r^*) = \epsilon V(r^*_{max})$, and defining $\Delta r^*= r^*-r^*_{max}$, we obtain $(\Delta r^*)^2 \approx -2(1-\epsilon)V(r^*_{max})/V^{(2)}$. Thus, the imaginary part of the QNFs to sub-leading order are given by
\begin{equation}
\omega_I^2= \frac{(1-\epsilon)}{4 (\Delta r^*)^2} -\frac{5 \mathcal{B}^2 (59+279 \Lambda M^2)}{209952 M^2 L^2}\,.
\end{equation}
Therefore, the width is found to behaves, to second order, as
\begin{equation}
(\Delta r^*)^2 \approx \frac{3(1-\epsilon) M^2}{\mathcal{B}} \left( 9+ \frac{1}{L^2}\mathcal{B} (-2 +135 m^2 M^2 -90 \Lambda M^2) \right)\,,
\end{equation}
and exhibit an inverted behavior. For $\Lambda=0$ the result of \cite{Lagos:2020oek} is recovered. We note that this inversion in the behaviour of the width does not happen exactly at the same critical value of the mass  where imaginary part of the QNFs exhibits an inverted behaviour, as in Ref. \cite{Lagos:2020oek}. However, certainly it is responsible of the inverted behaviour of $\omega_I$.

%, due to the presence of the second term in the RHS of eq. (33)

\section{Conclusions}
\label{conclusion}

In this work, we considered the Schwarzschild-dS and the Schwarzschild-AdS black hole as backgrounds and we studied the propagation of massive scalar fields through the QNFs by using the pseudospectral Chebyshev method in order to determine if there is an anomalous decay behaviour  in the QNMs as it was observed in the asymptotically flat Schwarzschild black hole background.

The QNMs of a massless scalar field in the background of a Schwarzschild-dS black hole are  characterized by two families of modes, the photon sphere modes 
which are complex and the dS modes
consisting of purely imaginary QNFs. The dS modes
are generated for small scalar field mass and eventually this family can acquires a real part when the mass of the scalar field increases, and it is worth to mention that
to our knowledge, this is the first time that this behaviour has been reported. Both families present
frequencies with a negative imaginary part, which means that the propagation of scalar field is stable in this background. 

For the photon sphere modes,
the presence of the cosmological constant leads to decrease the real oscillation frequency and to a slower decay \cite{Zhidenko:2003wq}. We showed that for the dominant QNFs there is a slower decay rate when the mass of the scalar field increases for a fixed angular harmonic number $\ell$. Furthermore, we showed the existence of anomalous decay rate of QNMs, i.e, the absolute values of the imaginary part of the QNFs decay when the angular harmonic numbers increase if the mass of the scalar field is smaller than a critical mass. On the contrary  they grow when the angular harmonic numbers increase, if the mass of the scalar field is larger than the critical mass and they also increase with the overtone number $n_{PS}$, for $\ell \geq n_{PS}$.
We also showed that the effect of the cosmological constant is to shift the values of the critical mass  i.e  the value of the critical mass increases with the cosmological constant. Moreover, by using the WKB method, we found that the critical mass for large $\ell$ is given by $m_c M = \sqrt{(137+18045 \Lambda M^2)/29160}$, for the dominant mode $n_{PS}=0$. It is worth to mention here that  the critical mass is an interesting quantity, because it shows that it is possible to have a scalar field with a critical mass and the decay rate does not depend  appreciably on the angular harmonic numbers $\ell$; however, its frequency of oscillation depends on the angular harmonic numbers $\ell$, increasing when $\ell$ increases.

For the dS modes,  it was shown,  depending on the black hole mass it may even be the dominant mode \cite{Jansen:2017oag}. We found that for a fixed value of the black hole mass, the purely imaginary QNFs can also be dominant depending on the scalar field mass and the angular harmonic numbers. Additionally, a faster decay is observed when the $\ell$ parameter increases, as well as, when the scalar field mass increases until that the QNFs acquire a real part, after it the decay is stabilized, and the frequency of the oscillations increases when the scalar field mass increases. Furthermore, we showed that this family does not present an anomalous behaviour of the QNFs, for the range of scalar field mass analyzed.

In the case of a  Schwarzschild-AdS black hole background we have shown that the QNMs  of massive scalar fields under Dirichlet boundary conditions present an anomalous behaviour of the QNMS because the longest-lived modes are always the ones with higher angular number. Also, they present a faster decay when the scalar field mass increases. However,  we showed that for large and intermediate black holes there is not a critical mass where the behaviour of the modes is inverted.  For small black holes our analysis was performed only for low values of $\ell$, suggesting that neither there is a critical mass in this case. However, it would be worthwhile to check this claim with some numerical method that works well for high $\ell$.

Therefore, the anomalous behaviour in the QNFs is possible in asymptotically flat, in asymptotically dS and
 in
 asymptotically AdS spacetimes; however, we observed that the critical mass exist for asymptotically flat and for asymptotically dS spacetimes and 
 it is not present in 
 asymptotically AdS spacetimes for large and intermediate black holes. Also, as we mentioned, through our analysis of low values of $\ell$ there is not indication that it could be present for small asymptotically AdS black holes.
 The existence of the critical mass
 could depend if the scalar field probes the divergence of the effective potential at infinity, despite that the boundary conditions can be imposed in a different point. It is worth to mention that for a Schwarzschild black hole, the effective potential tends to $m^2$, so the scalar field does not probe the divergence, and consequently a critical mass
 can be observed. Note that, for the geometries studied, the 
 critical mass
 is only present for the photon sphere modes, which are well described by the WKB approximation, and they approach asymptotically  to the QNMS of Schwarzschild black hole for very small value of the cosmological constant. On the other hand, the QNMs of Schwarzschild AdS black hole do not approach asymptotically  to the QNMS of Schwarzschild black hole for very small cosmological constant, due to the boundary conditions at spatial infinity and the effective potentials are very different.
 
It would be interesting to extent this work to the case the background black hole is charged and study the behaviour of QNMs in this background and in different asymptotic spacetimes.  If the background metric is the Reissner-Nordstr\"om  black hole in dS spacetime an interesting effect in Gravity theory and in its scalar-tensor extension, is  the Strong Cosmic Censorship (SCC) effect \cite{Cardoso:2017soq,Destounis:2019omd}. It was found in \cite{Cardoso:2018nvb} that there intervals of the masses and charges of the scalar field that we have violation or not of SCC. Therefore it would be interesting to study how this behaviour of SCC is connected with the anomalous decay of the QNMs and the critical mass of the scalar field beyond which these decays occur.

\appendix{}

\section{Accuracy of the numerical method}
\label{Accuracy}

In Table \ref{Table1} we show some dominant photon sphere modes,
in order to check the correctness and accuracy of the numerical technique used. Also, we show the relative error, which is defined by
\begin{equation}
\label{ERe}
\epsilon_{Re(\omega)} =\frac{\mid Re(\omega_1)- Re(\omega_0)\mid}{Re(\omega_0)}\cdot 100\%
\end{equation}
\begin{equation}
\label{EIm}
\epsilon_{Im(\omega)} =\frac{\mid Im(\omega_1)- Im(\omega_0)\mid}{Im(\omega_0)} \cdot 100\%
\end{equation}
where $\omega_1$ corresponds to the result from \cite{Zhidenko:2003wq}, and $\omega_0$ denotes our result.  The complex QNFs for this geometry was determined in Ref. \cite{Zhidenko:2003wq} by using the WKB and P\"oschl-Teller method. We can observed that error does not exceed $0.37\%$ when we compare our results with the WKB method and $2.198\%$ with the P-T method. As it was observed, the frequencies all have a negative imaginary part, which means that the propagation of scalar field is stable in this background. Also, we observe that the presence of a bigger cosmological constant leads to decrease the real oscillation frequency and to a slower decay.

\begin{table}[ht]
\caption{Dominant photon sphere modes ($n_{PS}=0$)
for massless  scalar fields with $\ell=1$ in the background of Schwarzchild-de Sitter black holes. The values of $\omega_{WKB}$ and $\omega_{P-T}$ appear in Ref. \cite{Zhidenko:2003wq}.}
\label{Table1}\centering
\scalebox{0.6}{
\begin{tabular}{ | c | c | c | c | c | c |c | c |}
\hline
$\Lambda$ & $\omega_{WKB}$ & $\omega_{P-T}$ & $\omega $ &  $\epsilon_{Re(\omega)} (WKB) $ & $\epsilon_ {Im(\omega)} (WKB) $ & $\epsilon_{Re(\omega)} (P-T)$ & $\epsilon_{Im(\omega)} (P-T)$\\ \hline
%$0$ & $0.2929 - 0.0978i$ & $0.299 -0.101i$ &   $...$ & %$...$& $...$ \\
$0.02$ & $0.2603 - 0.0911i$ & $ 0.263 - 0.093i$ &
$0.2603 - 0.0910 i$ &  $0.000$ & $0.110$ &  $1.037$ & $2.198$\\
$0.04$ & $0.2247 - 0.0821i$ & $0.226-0.083i$ &   $0.2247 - 0.0821 i$ &  $0.045$ & $0.122$ &  $0.623$ & $1.220 $\\
$0.06$ & $0.1854 - 0.0701i$ & $0.187 - 0.071i$ &   $0.1854 - 0.0701 i$  & $0.054$ & $0.143$ &  $0.917$ & $1.429$\\
$0.08$ & $0.1404 - 0.0542i$ & $0.141 - 0.055i$ &  $0.1404 - 0.0540 i$  & $0.000$ & $0.370$ &  $0.427$ & $1.852$\\
$0.09$ & $0.11392 - 0.04397i$ & $0.1147 - 0.0443i$ &   $0.11400 - 0.04388 i$ &  $0.070$ & $0.205$ &  $0.614$ & $0.957$\\
$0.10$ & $0.08156 - 0.03121i$ & $0.0819 - 0.0315i$ &   $0.08159 - 0.03123 i$ &  $0.037$ & $0.064$ &  $0.380 $ & $0.865$\\
$0.11$ & $0.02549 - 0.00965i$ & $0.02550 - 0.00967i$ &   $0.02549 - 0.00965 i$  & $0.000$ & $0.000$ &  $0.039$ & $0.207$\\ \hline
\end{tabular}}
\end{table}

On the other hand, in \cite{Jansen:2017oag} another branch of purely imaginary QNFs was found for this geometry by using the pseudospectral Chebyshev method, with the metric expressed in Eddington-Finkelstein coordinates. Here, we have considered the coordinates given by the metric. Eq. (\ref{metric}) along with the change of variables $y=(r-r_H)/(r_{\Lambda}-r_H)$. 
Now, in order to check the correctness and accuracy of the numerical techniques used, we show the dominant dS modes
in Table \ref{Table2}, where the relative error vanishes.
As it was observed, the frequencies all are negative, which means that the propagation of scalar field is stable in this background. However, the presence of the cosmological constant leads to a fast decay, when it increases, that is, contrary to the complex QNFs. Also, it was shown that depending on the black hole mass may even be the dominant modes \cite{Jansen:2017oag}.
\begin{table}[ht]
\caption{Dominant dS modes
($n_{dS}=0$) for massless  scalar fields with $\ell=1$ in the background of Schwarzchild-de Sitter black holes with $M=1$. The values of
$\omega_I$  appear in Ref. \cite{Jansen:2017oag}.}
\label{Table2}\centering
\scalebox{0.7}{
\begin{tabular}{ | c | c | c |}
\hline
$\Lambda$ & $ \omega_I$  & $\omega$  \\ \hline
$0.02$  &
$-0.081565496i$ &  $-0.081565496 i$  \\
$0.04$ &  $-0.11524810i$ &    $-0.11524810 i
$ \\
$0.06$ &   $-0.14100253i$ &  $ -0.14100253 i
$ \\
$0.08$  &  $-0.16268011i$ &  $-0.16268011i $  \\
$0.09$ &   $-0.17249210i$ &  $ -0.17249210i$   \\
$0.10$ &   $-0.18177480i$ &  $-0.18177480 i $  \\
$0.11$ &   $-0.19057630i$ &  $-0.19057630 i$  \\\hline
\end{tabular}}
\end{table}

\newpage

\section{Numerical values. Schwarzschild-de Sitter black hole with $M^2\Lambda=0.04$}
\label{tables}

In this appendix we provide the numerical values of the QNFs. All complex QNFs come with their negative complex conjugate which we do not show.
In Table \ref{T1} and \ref{T2} we show for both families the dominant modes 
for massive scalar fields in the background of a Schwarzschild-de Sitter black hole, and in Table \ref{T3}, the numerical values  of the photon sphere mode for the first overtone number ($n_{PS}=1$).

\begin {table}[ht]
\caption {Dominant photon sphere modes ($n_{PS}=0$) for massive scalar fields with $\ell = 0, 1, 2, 10, 20,$ and $30$ in the background of a Schwarzschild-de Sitter black hole with %$M=1$, and 
$M^2\Lambda = 0.04$.}
\label {T1}\centering
\scalebox{0.7}{
\begin {tabular} { | c | c | c | c | c |}
\hline
$\ell$ & $mM = 0.02 $ & $mM = 0.03 $ & $mM = 0.04 $ & $mM = 0.05 $ \\\hline
$0$ &
$0.08221623 - 0.10123911 i$ &
$0.08246910 - 0.10107321 i$ &
$0.08281206 - 0.10083840 i$ &
$ 0.08323382 - 0.10053157 i$ \\
$1$ &
$0.22485188 - 0.08198069 i$ &
$0.22506071 - 0.08189246 i$ &
$0.22535335 - 0.08176899 i$ &
$0.22573006 - 0.08161032 i$ \\
$2$ &
$0.38089209 - 0.07873423 i$ &
$0.38102729 - 0.07870351 i$ &
$0.38121659 - 0.07866053 i$ &
$0.38145998 - 0.07860531 i$ \\
$10$ &
$1.61559336 - 0.07707753 i$ &
$1.61562631 - 0.07707585 i$ &
$1.61567245 - 0.07707350 i $ &
$1.61573177 - 0.07707047 i $ \\
$20$ &
$3.15567507 - 0.07700559 i$ &
$3.15569196 - 0.07700515 i $ &
$3.15571561 - 0.07700453 i$ &
$3.15574602 - 0.07700374 i $ \\
$30$ &
$4.69544173 - 0.07699158 i$ &
$4.69545309 - 0.07699138 i $ &
$4.69546899 - 0.07699110 i $ &
$4.69548943 - 0.07699074 i $ \\\hline
$\ell$ & $mM = 0.06 $ & $mM = 0.07 $ & $mM = 0.08 $ & $mM = 0.09 $ \\\hline
$0$ &
$0.08371955 - 0.10014667 i $ &
$0.08425049 - 0.09967430 i $ &
$0.08480354 - 0.09909918 i$ &
$0.08535077 - 0.09839688 i $ \\
$1$ &
$0.22619110 - 0.08141653 i $ &
$0.22673721 - 0.08118773 i $ &
$0.22736862 - 0.08092405 i $ &
$0.22808606 - 0.08062571 i $ \\
$2$ &
$0.38175751 - 0.07853780 i $ &
$0.38210917 - 0.07845832 i $ &
$0.38251490 - 0.07836665 i$ &
$0.38297502 - 0.07826293 i$ \\
$10$ &
$1.61580427 - 0.07706678 i $ &
$1.61588995 - 0.07706241 i $ &
$1.61598882 - 0.07705737 i$ &
$1.61610086 - 0.07705166 i $ \\
$20$ &
$3.15578319 - 0.07700277 i $ &
$3.15582711 - 0.07700163 i $ &
$3.15587779 - 0.07700031 i$ &
$3.15593522 - 0.07699881 i $ \\
$30$ &
$4.69551441 - 0.07699031 i $ &
$4.69554394 - 0.07698979 i$ &
$4.69557801 - 0.07698919 i$ &
$4.69561662 - 0.07698852 i $ \\\hline
$\ell$ & $mM = 0.10 $ & $mM = 0.11 $ & $mM = 0.12 $ & $mM = 0.13 $ \\\hline
$0$ &
$0.08585920 - 0.09752810 i$ &
$0.08629220 - 0.09642921 i$ &
$0.08661627 - 0.09499337 i $ &
$0.08682704 - 0.09304079 i $ \\
$1$ &
$0.22889025 - 0.08029297 i$ &
$0.22978202 - 0.07992618 i $ &
$0.23076229 - 0.07952580 i $ &
$0.23183206 - 0.07909243 i $ \\
$2$ &
$0.38348926 - 0.07814724 i$ &
$0.38405777 - 0.07801965 i $ &
$0.38468056 - 0.07788026 i $ &
$0.38535769 - 0.07772916 i $ \\
$10$ &
$1.61622608 - 0.07704529 i$ &
$1.61636449 - 0.07703824 i $ &
$1.61651607 - 0.07703052 i $ &
$1.61668084 - 0.07702213 i i $ \\
$20$ &
$3.15599942 - 0.07699714 i$ &
$3.15607037 - 0.07699530 i $ &
$3.15614808 - 0.07699327 i$ &
$3.15623254 - 0.07699107 i $ \\
$30$ &
$4.69565977 - 0.07698776 i$ &
$4.69570747 - 0.07698693 i $ &
$4.69575971 - 0.07698601 i $ &
$4.69581649 - 0.07698502 i $ \\\hline
$\ell$ & $mM = 0.14 $ & $mM = 0.15 $ & $mM = 0.16 $ & $mM = 0.17 $ \\\hline
$0$ &
$0.08703483 - 0.09028329 i $ &
$0.08767928 - 0.08640066 i $ &
$0.08961851 - 0.08156211 i$ &
$0.09323258 - 0.07677868 i$ \\
$1$ &
$0.23299242 - 0.07862680 i $ &
$0.23424454 - 0.07812982 i $ &
$0.23558962 - 0.07760263 i$ &
$0.23702888 - 0.07704656 i $ \\
$2$ &
$0.38608920 - 0.07756646 i $ &
$0.38687512 - 0.07739228 i$ &
$0.38771552 - 0.07720675 i$ &
$0.38861043 - 0.07701000 i $ \\
$10$ &
$1.61685878 - 0.07701307 i $ &
$1.61704990 - 0.07700334 i $ &
$1.61725420 - 0.07699294 i$ &
$1.61747167 - 0.07698188 i$ \\
$20$ &
$3.15632376 - 0.07698870 i $ &
$3.15642174 - 0.07698615 i $ &
$3.15652647 - 0.07698342 i$ &
$3.15663796 - 0.07698052 i $ \\
$30$ &
$4.69587781 - 0.07698395 i $ &
$4.69594368 - 0.07698280 i $ &
$4.69601408 - 0.07698157 i$ &
$4.69608903 - 0.07698025 i $ \\\hline
$\ell$ & $mM = 0.18 $ & $mM = 0.19 $ & $mM = 0.20 $ & $mM = 0.21 $ \\\hline
$0$ &
$0.09793849 - 0.07278715 i$ &
$0.10312414 - 0.06961851 i$ &
$0.10848564 - 0.06709019 i $ &
$0.11389788 - 0.06503633 i$ \\
$1$ &
$0.23856353 - 0.07646322 i$ &
$0.24019471 - 0.07585443 i$ &
$0.24192347 - 0.07522232 i $ &
$0.24375068 - 0.07456924 i$ \\
$2$ &
$0.38955991 - 0.07680221 i$ &
$0.39056402 - 0.07658352 i$ &
$0.39162279 - 0.07635413 i $ &
$0.39273630 - 0.07611422 i$ \\
$10$ &
$1.61770232 - 0.07697014 i$ &
$1.61794614 - 0.07695775 i $ &
$1.61820314 - 0.07694468 i $ &
$1.61847332 - 0.07693095 i $ \\
$20$ &
$3.15675621 - 0.07697744 i$ &
$3.15688122 - 0.07697419 i $ &
$3.15701298 - 0.07697076 i $ &
$3.15715149 - 0.07696716 i $ \\
$30$ &
$4.69616853 - 0.07697886 i$ &
$4.69625256 - 0.07697739 i$ &
$4.69634114 - 0.07697585 i $ &
$4.69643426 - 0.07697422 i $ \\\hline
$\ell$ & $mM = 0.22 $ & $mM = 0.23 $ & $mM = 0.24 $ & $mM = 0.25 $ \\\hline
$0$ &
$0.11931155 - 0.06333786 i $ &
$0.12470831 - 0.06191201 i $ &
$0.13008281 - 0.06070044 i $ &
$0.13543506 - 0.05966101 i $ \\
$1$ &
$0.24567703 - 0.07389780 i $ &
$0.24770294 - 0.07321081 i $ &
$0.24982855 - 0.07251125 i$ &
$0.25205365 - 0.07180223 i $ \\
$2$ &
$0.39390459 - 0.07586401 i $ &
$0.39512771 - 0.07560372 i $ &
$0.39640571 - 0.07533360 i$ &
$0.39773863 - 0.07505389 i $ \\
$10$ &
$1.61875666 - 0.07691655 i $ &
$1.61905318 - 0.07690149 i $ &
$1.61936287 - 0.07688577 i$ &
$1.61968573 - 0.07686938 i $ \\
$20$ &
$3.15729676 - 0.07696338 i $ &
$3.15744879 - 0.07695942 i $ &
$3.15760757 - 0.07695529 i$ &
$3.15777311 - 0.07695099 i $ \\
$30$ &
$4.69653192 - 0.07697251 i $ &
$4.69663412 - 0.07697072 i $ &
$4.69674086 - 0.07696886  i$ &
$4.69685215 - 0.07696691 i $ \\\hline
$\ell$ & $mM = 0.26 $ & $mM = 0.28 $ & $mM = 0.30 $ & $mM = 0.32 $ \\\hline
$0$ &
$0.14076723 - 0.05876242 i$ &
$0.15138316 - 0.05729793 i$ &
$0.16195355 - 0.05617189 i$ &
$0.17249742 - 0.05529719 i$ \\
$1$ &
$0.25437769 - 0.07108691 i$ &
$0.25931866 - 0.06965002 i$ &
$0.26464006 - 0.06822498 i$ &
$0.27032567 - 0.06683365 i $ \\
$2$ &
$0.39912653 - 0.07476488 i$ &
$0.40206736 - 0.07416012 i$ &
$0.40522840 - 0.07352189 i$ &
$0.40860970 - 0.07285303 i$ \\
$10$ &
$1.62002176 - 0.07685233 i $ &
$1.62073331 - 0.07681625 i$ &
$1.62149753 - 0.07677753 i$ &
$1.62231440 - 0.07673618 i$ \\
$20$ &
$3.15794541 - 0.07694650 i$ &
$3.15831026 - 0.07693702 i$ &
$3.15870213 - 0.07692683 i$ &
$3.15912103 - 0.07691594 i$ \\
$30$ &
$4.69696798 - 0.07696489 i$ &
$4.69721326 - 0.07696060 i$ &
$4.69747671 - 0.07695599 i$ &
$4.69775833 - 0.07695107 i$ \\\hline
$\ell$ & $mM = 0.34 $ & $mM = 0.36 $ & $mM = 0.38 $ & $mM = 0.40 $ \\\hline
$0$ &
$0.18302899 - 0.05461419 i$ &
$0.19355845 - 0.05408027 i$ &
$0.20409290 - 0.05366384 i$ &
$0.21463710 - 0.05334089 i$ \\
$1$ &
$0.27635558 - 0.06549424 i$ &
$0.28270766 - 0.06422074 i$ &
$0.28935890 - 0.06302293 i$ &
$0.29628643 - 0.06190680 i$ \\
$2$ &
$0.41221109 - 0.07215671 i$ &
$0.41603215 - 0.07143638 i$ &
$0.42007211 - 0.07069577 i$ &
$0.42432983 - 0.06993871 i$ \\
$10$ &
$1.62318391 - 0.07669220 i$ &
$1.62410605 - 0.07664561 i$ &
$1.62508081 - 0.07659641 i$ &
$1.62610817 - 0.07654462 i$ \\
$20$ &
$3.15956693 - 0.07690435 i$ &
$3.16003986 - 0.07689206 i$ &
$3.16053979 - 0.07687908 i$ &
$3.16106674 - 0.07686540 i$ \\
$30$ &
$4.69805811 - 0.07694583 i$ &
$4.69837606 - 0.07694027 i$ &
$4.69871217 - 0.07693440 i$ &
$4.69906645 - 0.07692821 i$ \\\hline
\end {tabular} }
\end {table}

\begin {table}[ht]
\caption {Dominant dS modes ($n_{dS}=0$) for massive scalar fields with $\ell = 0, 1, 2, 10, 20,$ and $30$ in the background of a Schwarzschild-de Sitter black hole with %$M=1$, and 
$M^2\Lambda = 0.04$.}
\label {T2}
\centering
\scalebox{0.7}{
\begin {tabular} { | c | c | c | c | c |}
\hline
$\ell$ & $mM = 0.02 $ & $mM = 0.03 $ & $mM = 0.04 $ & $mM = 0.05 $ \\\hline
$0$ &
$-0.00089412 i$ &
$-0.00202588 i$ &
$-0.00363772 i$ &
$-0.00575958 i$ \\
$1$ &
$-0.11631962 i$ &
$-0.11767119 i$ &
$-0.11958675 i$ &
$-0.12209126 i$ \\
$2$ &
$-0.23192813 i$ &
$-0.23329938 i$ &
$-0.23524197 i$ &
$-0.23778024 i$
%\\
%$10$ &
%$-1.155803815 I$ &
%$-1.157218314 I$ &
%$-1.159220519 I$ &
%$-1.161833776 I $ \\
%$20$ &
%$-2.31052410 I$ &
%$-2.31195117 I $ &
%$-2.31397076 I$ &
%$-2.31660599 I $ \\
%$30$ &
%$-3.46441436 I$ &
%$-3.46581998 I $ &
%$-3.46780771 I $ &
%$-3.47039865 I $
\\\hline
$\ell$ & $mM = 0.06 $ & $mM = 0.07 $ & $mM = 0.08 $ & $mM = 0.09 $ \\\hline
$0$ &
$-0.00843402 i $ &
$-0.01172035 i $ &
$-0.01570111 i$ &
$-0.02049265 i $ \\
$1$ &
$-0.12521929 i$ &
$-0.12901740 i $ &
$-0.13354774 i $ &
$-0.13889355 i $ \\
$2$ &
$-0.24094791 i $ &
$-0.24479034 i $ &
$-0.24936799 i$ &
$-0.25476169 i$
%\\
%$10$ &
%$-1.165090368 I $ &
%$-1.169033645 I $ &
%$-1.173721246 I$ &
%$-1.179230044 I $ \\
%$20$ &
%$-2.31988880 I $ &
%$-2.32386206 I $ &
%$-2.32858274 I$ &
%$-2.33412673 I $ \\
%$30$ &
%$-3.47362189 I$ &
%$-3.47751640 I$ &
%$-3.48213383 I$ &
%$-3.48754317 I $
\\\hline
$\ell$ & $mM = 0.10 $ & $mM = 0.11 $ & $mM = 0.12 $ & $mM = 0.13 $ \\\hline
$0$ &
$-0.02626288 i$ &
$-0.03326237 i$ &
$-0.04188062 i$ &
$-0.05274799 i $ \\
$1$ &
$-0.14516806 i$ &
$-0.15252930 i$ &
$-0.16120646 i $ &
$-0.17155137 i $ \\
$2$ &
$-0.26108116 i$ &
$-0.26847912 i $ &
$-0.27717645 i $ &
$-0.28751105 i $
%\\
%$10$ &
%$-1.185663994 I$ &
%$-1.193167146 I $ &
%$-1.201946634 I $ &
%$-1.212316856 I $ \\
%$20$ &
%$-2.34059654 I$ &
%$-2.34813405 I$ &
%$-2.35694284 I$ &
%$-2.36733121 I $ \\
%$30$ &
%$-3.49383822 I$ &
%$-3.50115126 I $ &
%$-3.50968006 I $ &
%$-3.51975056 I $ \\\hline
\\\hline
$\ell$ & $mM = 0.14 $ & $mM = 0.15 $ & $mM = 0.16 $ & $mM = 0.17 $ \\\hline
$0$ &
$-0.06689633 i$ &
$-0.08585232 i $ &
$-0.11152914 i$ &
$-0.15564561 i$ \\
$1$ &
$-0.18415171 i $ &
$-0.20013275 i $ &
$-0.22228398 i$ &
$-0.26715306 i $ \\
$2$ &
$-0.30004513 i $ &
$-0.31584894 i$ &
$-0.33754487 i$ &
$-0.37950874 i $
%\\
%$10$ &
%$-1.224795094 I $ &
%$-1.240349041 I $ &
%$-1.261270881 I$ &
%$-1.298250328 I$ \\
%$20$ &
%$-2.37980465 I $ &
%$-2.39530405 I $ &
%$-2.41603754 I$ &
%$-2.45195589 I $ \\
%$30$ &
%%$-3.53200107 I $ &
%%$-3.54823828 I $ &
%%$0.01383577 - 3.57541559 I$ &
%$0.02992128 - 3.58111144 I $ \\\hline
\\\hline
$\ell$ & $mM = 0.18 $ & $mM = 0.19 $ & $mM = 0.20 $ & $mM = 0.21 $ \\\hline
$0$ &
$0.05353197 - 0.17154859 i$ &
$0.07353052 - 0.16967135 i$ &
$0.08727790 - 0.16851920 i$ &
$0.09823954 - 0.16766383 i$ \\
$1$ &
$0.06336411 - 0.28338941 i$ &
$0.09146811 - 0.27990619 i$ &
$0.11287548 - 0.27722512 i$ &
$0.13110568 - 0.27524716 i$ \\
$2$ &
$0.06076188 - 0.39913464 i$ &
$0.08893906 - 0.39604938 i$ &
$0.11043784 - 0.39339544 i $ &
$0.12880897 - 0.39098624 i$
%\\
%$10$ &
%$0.054953060 - 1.324880692 I$ &
%$0.083330087 - 1.323337101 I$ &
%$0.104871948 - 1.322009000 I$ &
%$0.123294803 - 1.320808005 I$ \\
%$20$ &
%$0.05307644 - 2.48016379 I$ &
%$0.08166609 - 2.47932643 I $ &
%$0.10340496 - 2.47841999 I $ &
%$0.12196939 - 2.47752623 I $ \\
%$30$ &
%$0.04021448 - 3.58725804 I$ &
%$0.04836959 - 3.59382049 I$ &
%$0.05515866 - 3.60072342 I $ &
%$0.06089046 - 3.60785393 I $ \\\hline
\\\hline
$\ell$ & $mM = 0.22 $ & $mM = 0.23 $ & $mM = 0.24 $ & $mM = 0.25 $ \\\hline
$0$ &
$0.10763027 - 0.16695588 i $ &
$0.11602271 - 0.16633358 i$ &
$0.12373171 - 0.16576818 i $ &
$0.13094823 - 0.16524502 i $ \\
$1$ &
$0.14737955 - 0.27403938 i $ &
$0.16214487 - 0.27374020 i $ &
$0.17539633 - 0.27438833 i$ &
$0.18697588 - 0.27570348 i$ \\
$2$ &
$0.14531497 - 0.38872888 i $ &
$0.16056963 - 0.38657115 i $ &
$0.17492361 - 0.38448131 i$ &
$0.18859932 - 0.38243893 i $
%\\
%$10$ &
%$1.61875666 - 0.07691655 i $ &
%$1.61905318 - 0.07690149 i $ &
%$1.61936287 - 0.07688577 i$ &
%$1.61968573 - 0.07686938 i $ \\
%$20$ &
%$3.15729676 - 0.07696338 i $ &
%$3.15744879 - 0.07695942 i $ &
%$3.15760757 - 0.07695529 i$ &
%$3.15777311 - 0.07695099 i $ \\
%$30$ &
%$4.69653192 - 0.07697251 i $ &
%$4.69663412 - 0.07697072 i $ &
%$4.69674086 - 0.07696886  i$ &
%$4.69685215 - 0.07696691 i $ \
\\\hline
$\ell$ & $mM = 0.26 $ & $mM = 0.28 $ & $mM = 0.30 $ & $mM = 0.32 $ \\\hline
$0$ &
$0.13779637 - 0.16475606 i$ &
$0.15070260 - 0.16386364 i$ &
$0.16288342 - 0.16306984 i$ &
$0.17458166 - 0.16236395 i$ \\
$1$ &
$0.19696460 - 0.27719964 i $ &
$0.21350983 - 0.27958773 i$ &
$0.22735961 - 0.28088060 i$ &
$0.23974155 - 0.28135785 i $ \\
$2$ &
$0.20174945 - 0.38043017 i$ &
$0.22689872 - 0.37647517 i$ &
$0.25104980 - 0.37254550 i$ &
$0.27481104 - 0.36851568 i$
%\\
%$10$ &
%$1.62002176 - 0.07685233 i $ &
%$1.62073331 - 0.07681625 i$ &
%$1.62149753 - 0.07677753 i$ &
%$1.62231440 - 0.07673618 i$ \\
%$20$ &
%$3.15794541 - 0.07694650 i$ &
%$3.15831026 - 0.07693702 i$ &
%$3.15870213 - 0.07692683 i$ &
%$3.15912103 - 0.07691594 i$ \\
%$30$ &
%$4.69696798 - 0.07696489 i$ &
%$4.69721326 - 0.07696060 i$ &
%$4.69747671 - 0.07695599 i$ &
%$4.69775833 - 0.07695107 i$ \\\hline
\\\hline
$\ell$ & $mM = 0.34 $ & $mM = 0.36 $ & $mM = 0.38 $ & $mM = 0.40 $ \\\hline
$0$ &
$0.18594576 - 0.16173872 i$ &
$0.19707225 - 0.16118815 i$ &
$0.20802686 - 0.16070674 i$ &
$0.21885598 - 0.16028920 i$ \\
$1$ &
$0.25124153 - 0.28130544 i$ &
$0.26217041 - 0.28091663 i$ &
$0.27271023 - 0.28031643 i$ &
$0.28297559 - 0.27958645 i$ \\
$2$ &
$0.29906539 - 0.36392925 i$ &
$0.32509244 - 0.35591310 i$ &
$0.34460718 - 0.34125596 i$ &
$0.35774257 - 0.33038504 i$ \\\hline
%$10$ &
%$1.62318391 - 0.07669220 i$ &
%$1.62410605 - 0.07664561 i$ &
%$1.62508081 - 0.07659641 i$ &
%$1.62610817 - 0.07654462 i$ \\
%$20$ &
%$3.15956693 - 0.07690435 i$ &
%$3.16003986 - 0.07689206 i$ &
%$3.16053979 - 0.07687908 i$ &
%$3.16106674 - 0.07686540 i$ \\
%$30$ &
%$4.69805811 - 0.07694583 i$ &
%$4.69837606 - 0.07694027 i$ &
%$4.69871217 - 0.07693440 i$ &
%$4.69906645 - 0.07692821 i$ \\\hline
\end {tabular} }
\end {table}

\begin {table}[ht]
\caption {Photon sphere modes for the first overtone number ($n_{PS} =
      1 $) for massive scalar fields with $\ell = 1, 2, 10, 20,$ and $30$ in the background of a Schwarzschild-de Sitter black hole with 
      %$M=1$, and 
      $M^2\Lambda = 0.04$.}
\label {T3}\centering
\scalebox{0.7}{
\begin {tabular} { | c | c | c | c | c |}
\hline
$\ell$ & $mM = 0.02 $ & $mM = 0.03 $ & $mM = 0.04 $ & $mM = 0.05 $ \\\hline
%$0$ &
%$0.07386873 - 0.27084563 i$ &
%$0.07306139 - 0.27018604 i$ &
%$0.07191314 - 0.26911513 i $ &
%$0.07046879 - 0.26743660 i $ \\
$1$ &
$0.21538154 - 0.24734336 i$ &
$0.21546087 - 0.24716801 i$ &
$0.21557160 - 0.24692237 i $ &
$0.21571337 - 0.24660629 i $ \\
$2$ &
$0.37173556 - 0.23788191 i$ &
$0.37183876 - 0.23779619 i $ &
$0.37198332 - 0.23767614 i $ &
$0.37216932 - 0.23752171 i $ \\
$10$ &
$1.61300608 - 0.23133660 i$ &
$1.61303869 - 0.23133156 i$ &
$1.61308436 - 0.23132451 i $ &
$1.61314307 - 0.23131544 i $ \\
$20$ &
$3.15434249 - 0.23104400 i$ &
$3.15435934 - 0.23104268 i $ &
$3.15438293 - 0.23104083 i $ &
$3.15441326 - 0.23103845 i $ \\
$30$ &
$4.69454512 - 0.23098703 i$ &
$4.69455646 - 0.23098643 i $ &
$4.69457234 - 0.23098560 i $ &
$4.69459276 - 0.23098453 i $ \\\hline
$\ell$ & $mM = 0.06 $ & $mM = 0.07 $ & $mM = 0.08 $ & $mM = 0.09 $ \\\hline
%$0$ &
%$0.06896079 - 0.26488171 i $ &
%$0.06796546 - 0.26137874 i$ &
%$0.06800875 - 0.25749113 i$ &
%$0.06891396 - 0.25395753 i $ \\
$1$ &
$0.21588561 - 0.24621954 i $ &
$0.21608750 - 0.24576184 i$ &
$0.21631786 - 0.24523281 i$ &
$0.21657503 - 0.24463189 i $ \\
$2$ &
$0.37239683 - 0.23733284 i $ &
$0.37266599 - 0.23710945 i $ &
$0.37297694 - 0.23685145 i$ &
$0.37332984 - 0.23655872 i $ \\
$10$ &
$1.61321482 - 0.23130435 i $ &
$1.61329962 - 0.23129125 i $ &
$1.61339747 - 0.23127614 i$ &
$1.61350837 - 0.23125901 i $ \\
$20$ &
$3.15445032 - 0.23103555 i $ &
$3.15449413 - 0.23103212 i $ &
$3.15454468 - 0.23102816 i$ &
$3.15460196 - 0.23102367 i $ \\
$30$ &
$4.69461771 - 0.23098321 i $ &
$4.69464720 - 0.23098166 i $ &
$4.69468123 - 0.23097988 i $ &
$4.69471979 - 0.23097785 i $ \\\hline
$\ell$ & $mM = 0.10 $ & $mM = 0.11 $ & $mM = 0.12 $ & $mM = 0.13 $ \\\hline
%$0$ &
%$0.07018286 - 0.25106660 i$ &
%$0.07142593 - 0.24883952 i $ &
%$0.07235419 - 0.24727675 i $ &
%$0.07267182 - 0.24647570 i $ \\
$1$ &
$0.21685673 - 0.24395828 i$ &
$0.21715983 - 0.24321078 i $ &
$0.21748012 - 0.24238754 i $ &
$0.21781201 - 0.24148573 i $ \\
$2$ &
$0.37372489 - 0.23623115 i$ &
$0.37416230 - 0.23586859 i $ &
$0.37464233 - 0.23547088 i $ &
$0.37516526 - 0.23503782 i $ \\
$10$ &
$1.61363231 - 0.23123987 i$ &
$1.61376929 - 0.23121872 i $ &
$1.61391933 - 0.23119555 i $ &
$1.61408240 - 0.23117038 i $ \\
$20$ &
$3.15466598 - 0.23101866 i$ &
$3.15473675 - 0.23101312 i $ &
$3.15481425 - 0.23100705 i $ &
$3.15489849 - 0.23100045 i $ \\
$30$ &
$4.69476290 - 0.23097559 i$ &
$4.69481053 - 0.23097308 i$ &
$4.69486271 - 0.23097034 i $ &
$4.69491942 - 0.23096736 i $ \\\hline
$\ell$ & $mM = 0.14 $ & $mM = 0.15 $ & $mM = 0.16 $ & $mM = 0.17 $ \\\hline
%$0$ &
%$0.07194594 - 0.24679374 i $ &
%$0.06940721 - 0.24951312 i$ &
%$0.06665376 - 0.26100882 i$ &
%$0.07899327 - 0.27185380 i$ \\
$1$ &
$0.21814812 - 0.24050097 i $ &
$0.21847889 - 0.23942645 i $ &
$0.21879209 - 0.23825164 i$ &
$0.21907242 - 0.23696008 i $ \\
$2$ &
$0.37573141 - 0.23456922 i $ &
$0.37634114 - 0.23406483 i $ &
$0.37699485 - 0.23352441 i$ &
$0.37769299 - 0.23294766 i$ \\
$10$ &
$1.61425853 - 0.23114320 i $ &
$1.61444770 - 0.23111401 i $ &
$1.61464991 - 0.23108282 i$ &
$1.61486517 - 0.23104962 i $ \\
$20$ &
$3.15498947 - 0.23099333 i $ &
$3.15508719 - 0.23098567 i $ &
$3.15519164 - 0.23097749 i$ &
$3.15530284 - 0.23096879 i $ \\
$30$ &
$4.69498067 - 0.23096414 i $ &
$4.69504646 - 0.23096069 i $ &
$4.69511678 - 0.23095699 i$ &
$4.69519164 - 0.23095306 i $ \\\hline
$\ell$ & $mM = 0.18 $ & $mM = 0.19 $ & $mM = 0.20 $ & $mM = 0.21 $ \\\hline
%$0$ &
%$0.09005110 - 0.27444415 i$ &
%$0.09919617 - 0.27535494 i $ &
%$0.10731753 - 0.27566955 i$ &
%$0.11480745 - 0.27570066 i $ \\
$1$ &
$0.21930139 - 0.23552596 i$ &
$0.21945849 - 0.23390833 i $ &
$0.21952555 - 0.23204175 i $ &
$0.21950065 - 0.22982255 i $ \\
$2$ &
$0.37843607 - 0.23233427 i$ &
$0.37922468 - 0.23168389 i $ &
$0.38005945 - 0.23099615 i $ &
$0.38094113 - 0.23027065 i $ \\
$10$ &
$1.61509348 - 0.23101441 i$ &
$1.61533483 - 0.23097721 i $ &
$1.61558923 - 0.23093800 i $ &
$1.61585667 - 0.23089680 i$ \\
$20$ &
$3.15542078 - 0.23095956 i$ &
$3.15554545 - 0.23094980 i $ &
$3.15567686 - 0.23093951 i $ &
$3.15581501 - 0.23092870 i $ \\
$30$ &
$4.69527104 - 0.23094889 i$ &
$4.69535498 - 0.23094448 i $ &
$4.69544345 - 0.23093983 i $ &
$4.69553646 - 0.23093495 i $ \\\hline
$\ell$ & $mM = 0.22 $ & $mM = 0.23 $ & $mM = 0.24 $ & $mM = 0.25 $ \\\hline
%$0$ &
%$0.12186713 - 0.27557615 i $ &
%$0.12861336 - 0.27535931 i $ &
%$0.13512025 - 0.27508543 i$ &
%$0.14143801 - 0.27477583 i$ \\
$1$ &
$0.21943499 - 0.22709642 i $ &
$0.21950888 - 0.22368309 i $ &
$0.22009313 - 0.21952119 i$ &
$0.22158161 - 0.21487649 i$ \\
$2$ &
$0.38187055 - 0.22950693 i $ &
$0.38284868 - 0.22870456 i $ &
$0.38387659 - 0.22786306 i$ &
$0.38495554 - 0.22698194 i $ \\
$10$ &%
$1.61613716 - 0.23085359 i $ &
$1.61643069 - 0.23080840 i $ &
$1.61673726 - 0.23076121 i$ &
$1.61705688 - 0.23071203 i $ \\
$20$ &
$3.15595990 - 0.23091736 i $ &
$3.15611153 - 0.23090549 i $ &
$3.15626989 - 0.23089310 i$ &
$3.15643499 - 0.23088018 i $ \\
$30$ &%
$4.69563400 - 0.23092982 i$ &
$4.69573608 - 0.23092446 i $ &
$4.69584270 - 0.23091886 i$ &
$4.69595385 - 0.23091303 i $
\\\hline
$\ell$ & $mM = 0.26 $ & $mM = 0.28 $ & $mM = 0.30 $ & $mM = 0.32 $ \\\hline
%$0$ &
%$0.14760242 - 0.27444418 i$ &
%$0.15957099 - 0.27374860 i$ &
%$0.17117380 - 0.27304372 i$ &
%$0.18250487 - 0.27235298 i$ \\
$1$ &
$0.22405623 - 0.21022375 i$ &
$0.23106721 - 0.20200213 i$ &
$0.23944221 - 0.19545365 i$ &
$0.24834770 - 0.19022913 i$ \\
$2$ &
$0.38608693 - 0.22606075 i$ &
$0.38851376 - 0.22409645 i$ &
$0.39117274 - 0.22196770 i$ &
$0.39408350 - 0.21967447 i$ \\
$10$ &
$1.61738954 - 0.23066086 i$ &
$1.61809400 - 0.23055256 i$ &
$1.61885062 - 0.23043634 i$ &
$1.61965942 - 0.23031221 i$ \\
$20$ &
$3.15660683 - 0.23086673 i$ &
$3.15697073 - 0.23083827 i$ &
$3.15736157 - 0.23080770 i$ &
$3.15777937 - 0.23077503 i$ \\
$30$ &
$4.69606954 - 0.23090695 i$ &
$4.69631453 - 0.23089408 i$ &
$4.69657767 - 0.23088027 i$ &
$4.69685895 - 0.23086550 i$ \\\hline
$\ell$ & $mM = 0.34 $ & $mM = 0.36 $ & $mM = 0.38 $ & $mM = 0.40 $ \\\hline
%$0$ &
%$0.19362836 - 0.27168938 i$ &
%$0.20459029 - 0.27106026 i$ &
%$0.21542498 - 0.27046979 i$ &
%$0.22615872 - 0.26992011 i$ \\
$1$ &
$0.25745578 - 0.18597950 i$ &
$0.26663309 - 0.18245622 i$ &
$0.27582463 - 0.17948705 i$ &
$0.28500951 - 0.17695060 i$ \\
$2$ &
$0.39726960 - 0.21722114 i$ &
$0.40075749 - 0.21461877 i$ &
$0.40457407 - 0.21188714 i$ &
$0.40874302 - 0.20905562 i$\\
$10$ &
$1.62052039 - 0.23018019 i$ &
$1.62143352 - 0.23004031 i$ &
$1.62239881 - 0.22989260 i$ &
$1.62341626 - 0.22973707 i$ \\
$20$ &
$3.15822411 - 0.23074026 i$ &
$3.15869579 - 0.23070340 i$ &
$3.15919442 - 0.23066444 i$ &
$3.15972000 - 0.23062339 i$ \\
$30$ &
$4.69715838 - 0.23084978 i$ &
$4.69747595 - 0.23083311 i$ &
$4.69781166 - 0.23081549 i$ &
$4.69816552 - 0.23079692 i$
\\\hline
\end {tabular} }
\end {table}

\clearpage

\section{Numerical values. Schwarzschild-de Sitter black holes with $M^2\Lambda=0.11$}
\label{Tables2}
\begin {table}[ht]
\caption {Fundamental quasinormal frequencies ($n=0$) for massive scalar fields with $\ell = 0, 1, 2, 10, 20,$ and $30$ in the background of a Schwarzschild-de Sitter black hole with %$M=1$, and 
$M^2\Lambda = 0.11$.}
\label {T6}\centering
\scalebox{0.65}{
\begin {tabular} { | c | c | c | c | c |}
\hline
$\ell$ & $mM = 0.02 $ & $mM = 0.03 $ & $mM = 0.04 $ & $mM = 0.05 $ \\\hline
$0$ &
$0.0020380076 - 0.0189811106 i$ &
$0.0020383581 - 0.0189800488 i$ &
$0.0020327260 - 0.0189784691 i$ &
$0.0020144687 - 0.0189762687 i$ \\
$1$ &
$0.02551665 - 0.00964921 i$ &
$0.02554935 - 0.00964894 i$ &
$0.02559507 - 0.00964856 i$ &
$0.02565373 - 0.00964808 i$ \\
$2$ &
$0.04618334 - 0.00963129 i$ &
$0.04620139 - 0.00963121 i$ &
$0.04622664 - 0.00963112 i$ &
$0.04625909 - 0.00963099 i$ \\
$10$ &
$0.20162233 - 0.00962298 i$ &
$0.20162647 - 0.00962298 i$ &
$0.20163225 - 0.00962297 i$ &
$0.20163969 - 0.00962297 i$ \\
$20$ &
$0.39429226 - 0.00962263 i$ &
$0.39429437 - 0.00962263 i $ &
$0.39429733 - 0.00962263 i$ &
$0.39430114 - 0.00962263 i$ \\
$30$ &
$0.58681792 - 0.00962256 i$ &
$0.58681934 - 0.00962256 i$ &
$0.58682133 - 0.00962256 i$ &
$0.58682389 - 0.00962256 i $ \\\hline
$\ell$ & $mM = 0.06 $ & $mM = 0.07 $ & $mM = 0.08 $ & $mM = 0.09 $ \\\hline
$0$ &
$0.0019739262 - 0.0189732995 i$ &
$0.0018966550 - 0.0189693514 i$ &
$0.0017591920 - 0.0189641247 i$ &
$0.0015166966 - 0.0189571824 i$ \\
$1$ &
$0.02572524 - 0.00964751 i$ &
$0.02580950 - 0.00964683 i$ &
$0.02590639 - 0.00964607 i$ &
$0.02601575 - 0.00964522 i$ \\
$2$ &
$0.04629873 - 0.00963084 i$ &
$0.04634552 - 0.00963066 i$ &
$0.04639946 - 0.00963045 i$ &
$0.04646051 - 0.00963022 i$ \\
$10$ &
$0.20164878 - 0.00962296 i$ &
$0.20165953 - 0.00962295 i$ &
$0.20167193 - 0.00962294 i$ &
$0.20168597 - 0.00962293 i$ \\
$20$ &
$0.39430579 - 0.00962262 i$ &
$0.39431128 - 0.00962262 i$ &
$0.39431762 - 0.00962262 i$ &
$0.39432481 - 0.00962262 i$ \\
$30$ &
$0.58682701 - 0.00962256 i$ &
$0.58683071 - 0.00962256 i$ &
$0.58683497 - 0.00962256 i$ &
$0.58683979 - 0.00962256 i$ \\\hline
$\ell$ & $mM = 0.10 $ & $mM = 0.11 $ & $mM = 0.12 $ & $mM = 0.13 $ \\\hline
$0$ &
$0.0010460297 - 0.0189478654 i$ &
$-0.0181039644 i$ &
$-0.0171016030 i $ &
$-0.0162275285 i $ \\
$1$ &
$0.02613743 - 0.00964429 i$ &
$0.02627127 - 0.00964329 i $ &
$0.02641707 - 0.00964226 i $ &
$0.02657464 - 0.00964110 i$ \\
$2$ &
$0.04652865 - 0.00962996 i$ &
$0.04660385 - 0.00962967 i$  &
$0.04668607 - 0.00962937 i$ &
$0.04677528 - 0.00962904 i $ \\
$10$ &
$0.20170168 - 0.00962292 i$ &
$0.20171903 - 0.00962290 i$ &
$0.20173803 - 0.00962289 i$ &
$0.20175868 - 0.00962287 i$ \\
$20$ &
$0.39433284 - 0.00962261 i$ &
$0.39434171 - 0.00962261 i$ &
$0.39435143 - 0.00962261 i$ &
$0.39436110 - 0.00962260 i$ \\
$30$ &
$0.58684519 - 0.00962255 i$ &
$0.58685115 - 0.00962255 i$ &
$0.58685769 - 0.00962255 i$ &
$0.58686479 - 0.00962255 i$ \\\hline
$\ell$ & $mM = 0.14 $ & $mM = 0.15 $ & $mM = 0.16 $ & $mM = 0.17 $ \\\hline
$0$ &
$-0.0152852132 i$ &
$-0.0141564333 i$ &
$-0.01258354445 i$ &
$0.00189593 - 0.00973749 i $ \\
$1$ &
$0.02674378 - 0.00963992 i $ &
$0.02692425 - 0.00963869 i $ &
$0.02711585 - 0.00963742 i $ &
$0.02731832 - 0.00963611 i $ \\
$2$ &
$0.04687144 - 0.00962868 i$ &
$0.04697450 - 0.00962831 i$ &
$0.04708443 - 0.00962791 i$ &
$0.04720116 - 0.00962750 i$ \\
$10$ &
$0.20178099 - 0.00962285 i$ &
$0.20180494 - 0.00962283 i$ &
$0.20183054 - 0.00962281 i$ &
$0.20185779 - 0.00962279 i$ \\
$20$ &
$0.39437341 - 0.00962260 i$ &
$0.39438567 - 0.00962259 i$ &
$0.39439877 - 0.00962259 i$ &
$0.39441271 - 0.00962258 i$ \\
$30$ &
$0.58687245 - 0.00962255 i$ &
$0.58688069 - 0.00962254 i$ &
$0.58688949 - 0.00962254 i$ &
$0.58683071 - 0.00962256 i$ \\\hline
$\ell$ & $mM = 0.18 $ & $mM = 0.19 $ & $mM = 0.20 $ & $mM = 0.21 $ \\\hline
$0$ &
$0.00396067 - 0.00971436 i$ &
$0.00532622 - 0.00969559 i$ &
$0.00645433 - 0.00968015 i$ &
$0.00745450 - 0.00966731 i$ \\
$1$ &
$0.02753144 - 0.00963478 i$ &
$0.02775495 - 0.00963343 i$ &
$0.02798862 - 0.00963206 i$ &
$0.02823218 - 0.00963068 i$ \\
$2$ &
$0.04732466 - 0.00962707 i$ &
$0.04745487 - 0.00962662 i$ &
$0.04759173 - 0.00962615 i$ &
$0.04773520 - 0.00962566 i$ \\
$10$ &
$0.20188669 - 0.00962277 i$ &
$0.20191723 - 0.00962274 i$ &
$0.20194942 - 0.00962272 i$ &
$0.20198326 - 0.00962269 i$ \\
$20$ &
$0.39442750 - 0.00962257 i$ &
$0.39444314 - 0.00962257 i$ &
$0.39445961 - 0.00962256 i$ &
$0.39447694 - 0.00962255 i$ \\
$30$ &
$0.58690880 - 0.00962254 i$ &
$0.58691931 - 0.00962253 i$ &
$0.58693039 - 0.00962253 i$ &
$0.58694203 - 0.00962253 i$ \\\hline
$\ell$ & $mM = 0.22 $ & $mM = 0.23 $ & $mM = 0.24 $ & $mM = 0.25 $ \\\hline
$0$ &
$0.00837329 - 0.00965655 i$ &
$0.00923542 - 0.00964746 i$ &
$0.01005577 - 0.00963973 i$ &
$0.01084402 - 0.00963313 i$ \\
$1$ &
$0.02848538 - 0.00962930 i$ &
$0.02874797 - 0.00962792 i$ &
$0.02901970 - 0.00962654 i$ &
$0.02930030 - 0.00962518 i$ \\
$2$ &
$0.04788520 - 0.00962517 i$ &
$0.04804168 - 0.00962465 i$ &
$0.04820458 - 0.00962413 i$ &
$0.04837383 - 0.00962360 i$ \\
$10$ &
$0.20201874 - 0.00962266 i$ &
$0.20205586 - 0.00962263 i$ &
$0.20209463 - 0.00962260 i$ &
$0.20213503 - 0.00962257 i$ \\
$20$ &
$0.39449510 - 0.00962255 i$ &
$0.39451412 - 0.00962254 i$ &
$0.39453397 - 0.00962253 i$ &
$0.39455467 - 0.00962252 i$ \\
$30$ &
$0.58695424 - 0.00962252 i$ &
$0.58696702 - 0.00962252 i$ &
$0.58698036 - 0.00962252 i$ &
$0.58699427 - 0.00962251 i$ \\\hline
$\ell$ & $mM = 0.26 $ & $mM = 0.28 $ & $mM = 0.30 $ & $mM = 0.32 $ \\\hline
$0$ &
$0.01160687 - 0.00962747 i$ &
$0.01307445 - 0.00961838 i$ &
$0.01448450 - 0.00961154 i$ &
$0.01585274 - 0.00960634 i$ \\
$1$ &
$0.02958954 - 0.00962383 i$ &
$0.03019290 - 0.00962118 i$ &
$0.03082781 - 0.00961862 i$ &
$0.03149238 - 0.00961618 i$ \\
$2$ &
$0.04854937 - 0.00962305 i$ &
$0.04891903 - 0.00962193 i$ &
$0.04931301 - 0.00962078 i$ &
$0.04973074 - 0.00961961 i$ \\
$10$ &
$0.20217708 - 0.00962254 i$ &
$0.20226610 - 0.00962247 i$ &
$0.20236167 - 0.00962239 i$ &
$0.20246377 - 0.00962231 i$ \\
$20$ &
$0.39457621 - 0.00962251 i$ &
$0.39462183 - 0.00962250 i$ &
$0.39467082 - 0.00962248 i$ &
$0.39472318 - 0.00962245 i$ \\
$30$ &
$0.58700875 - 0.00962251 i$ &
$0.58703942 - 0.00962250 i$ &
$0.58707235 - 0.00962249 i$ &
$0.58710755 - 0.00962248 i$ \\\hline
$\ell$ & $mM = 0.34 $ & $mM = 0.36 $ & $mM = 0.38 $ & $mM = 0.40 $ \\\hline
$0$ &
$0.01718939 - 0.00960235 i$ &
$0.01850145 - 0.00959927 i$ &
$0.01979391 - 0.00959688 i$ &
$0.02107044 - 0.00959501 i$ \\
$1$ &
$0.03218476 - 0.00961386 i$ &
$0.03290322 - 0.00961167 i$ &
$0.03364608 - 0.00960962 i$ &
$0.03441178 - 0.00960771 i$ \\
$2$ &
$0.05017161 - 0.00961843 i$ &
$0.05063505 - 0.00961724 i$ &
$0.05112043 - 0.00961605 i$ &
$0.05162713 - 0.00961487 i$ \\
$10$ &
$0.20257241 - 0.00962223 i$ &
$0.20268758 - 0.00962214 i$ &
$0.20280925 - 0.00962204 i$ &
$0.20293742 - 0.00962194 i$ \\
$20$ &
$0.39477891 - 0.00962243 i$ &
$0.39483802 - 0.00962241 i$ &
$0.39490049 - 0.00962238 i$ &
$0.39496632 - 0.00962236 i$ \\
$30$ &
$0.58714502 - 0.00962247 i$ &
$0.58718476 - 0.00962246 i$ &
$0.58722677 - 0.00962245 i$ &
$0.58727105 - 0.00962244 i$ \\\hline
\end {tabular} }
\end {table}

\clearpage

\section{Numerical values. Schwarzschild-AdS black holes }
%with $r_H/R=10$}
\label{C}
\begin{table}[ht]
\caption{Fundamental quasinormal frequencies $\omega M$ ($n=0$) of massive scalar fields with $\ell=0,1,2,10,20,30$ in the background of  Schwarzschild-AdS black holes with $M^2\Lambda =-765075$, or $r_H/R=10$, where $R$ is the AdS radius.}
\label{T7}\centering
\scalebox{0.7}{
\begin{tabular}{ | c | c | c | c | c | c | c |c |}
\hline
$\ell$ & $mM = 0 $ & $mM = 0.05 $ & $mM = 0.10 $ & $mM = 0.15 $ \\\hline
$0$ & $18.60702097 - 26.64177595 i $ & $18.61223099 - 26.65137963 i$ & $ 18.62784397 - 26.68016019 i $ & $18.65380900 - 26.72802663 i$  \\
$1$ & $18.68948190 - 26.61845039 i $ & $ 18.69467400 - 26.62805066 i $ & $ 18.71023335 - 26.65682099 i $ & $ 18.73610934 - 26.70467047 i$ \\
$2$ & $18.85273204 - 26.57247886 i$ & $ 18.85788929 - 26.58207228 i $ & $18.87334427 - 26.61082210 i $ & $ 18.89904696 - 26.65863754 i$ \\
$10$ & $22.51253455 - 25.61606706 i$ & $22.51707726 - 25.62548491 i $ & $ 22.53069166 - 25.65370904 i $ & $ 22.55333682 - 25.70065162 i$ \\
$20$ & $30.21903865 - 24.02953567 i$ & $ 30.22286541 - 24.03855067 i$ & $ 30.23433465 - 24.06556806 i $ & $ 30.25341348 - 24.11050556 i $ \\
$30$ & $39.09826401 - 22.71395892 i$ & $ 39.10165741 - 22.72256722 i $ & $ 39.11182789 - 22.74836610 i $ & $ 39.12874649 - 22.79127794 i $  \\\hline
$\ell$ & $mM = 0.20 $ & $mM = 0.25 $ & $mM = 0.30 $ & $mM = 0.35 $ \\\hline
$0$ &$18.69004218 - 26.79482897 i$ & $ 18.73642795 - 26.88036058 i$ & $18.79282102 - 26.98436146 i$ & $ 18.85904854 - 27.10652214 i$  \\
$1$ & $18.77221856 - 26.77144920 i$ & $ 18.81844615 - 26.85695073 i $ & $ 18.87464766 - 26.96091523 i $ & $ 18.94065129 - 27.08303345 i$ \\
$2$ & $18.93491491 - 26.72536893 i$ & $ 18.98083460 - 26.81081010 i$ & $ 19.03666324 - 26.91470159 i $ & $ 19.10223099 - 27.03673460 i $ \\
$10$ & $22.58494526 - 25.76616790 i$ & $ 22.62542404 - 25.85005843 i $ & $ 22.67465618 - 25.95207222 i $ & $ 22.73250244 - 26.07191052 i $ \\
$20$ &  $30.28004766 - 24.17322743 i$ & $ 30.31416248 - 24.25354675 i$ & $ 30.35566387 - 24.35122822 i $ & $ 30.40443990 - 24.46599175 i $ \\
$30$ & $39.15236549 - 22.85117479 i$ & $ 39.18261910 - 22.92788037 i$ & $39.21942456 - 23.02117280 i $ & $ 39.26268334 - 23.13078800 i $ \\\hline
$\ell$ & $mM = 0.40 $ & $mM = 0.45 $ & $mM = 0.50 $ & $mM = 0.60 $ \\\hline
$0$ & $18.93491276 - 27.24648824 i$ & $ 19.02019383 - 27.40386537 i $ & $19.11465281 - 27.57822445 i$ & $19.33007160 - 27.97603061 i$  \\
$1$ & $19.01626044 - 27.22295125 i $ & $19.10125653 - 27.38027456 i$ & $ 19.19540202 - 27.55457456 i$ & $19.41011414 - 27.95224815 i$ \\
$2$ & $19.17734349 - 27.17655550 i$ & $19.26178462 - 27.33377075 i $ & $ 19.35531951 - 27.50795215 i $ & $19.56865488 - 27.90535950 i$ \\
$10$ & $22.79880337 - 26.20923117 i$ & $ 22.87338152 - 26.36365336 i $ & $ 22.95604382 - 26.53476268 i $ & $23.14478480 - 26.92524821 i$ \\ 
$20$ & $30.46036235 - 24.59751656 i$ & $ 30.52328849 - 24.74544559 i$ & $ 30.59306299 - 24.90939025 i$ & $30.75248432 - 25.28364347 i$ \\
$30$ & $39.31228260 - 23.25642346 i$ & $39.36809675 - 23.39774250 i$ & $ 39.42998910 - 23.55437869 i $ & $39.57141652 - 23.91201550 i$ \\\hline
\end{tabular}}
\end{table}

\begin{table}[ht]
\caption{Fundamental quasinormal frequencies $\omega M$ ($n=0$) of massive scalar fields with $\ell=0,1,2,10,20,30$ in the background of  Schwarzschild-AdS black holes with $M^2\Lambda =-1.0$.}
\label{T72}\centering
\scalebox{0.7}{
\begin{tabular}{ | c | c | c | c | c | c | c |c |}
\hline
$\ell$ & $mM = 0 $ & $mM = 0.05 $ & $mM = 0.10 $ & $mM = 0.15 $ \\\hline
$0$ & 
$1.46486212 - 1.14108905 i$ & 
$1.46604026 - 1.14240948 i$ & 
$1.46956344 - 1.14635875 i$ & 
$1.47539846 - 1.15290143 i$  \\
$1$ & 
$1.81421829 - 1.01340851 i$ & 
$1.81524513 - 1.01466010 i$ & 
$1.81831636 - 1.01840398 i$ & 
$1.82340447 - 1.02460801 i$ \\
$2$ & 
$2.30384515 - 0.87938138 i$ & 
$2.30476262 - 0.88053246 i$ & 
$2.30750672 - 0.88397613 i$ & 
$2.31205302 - 0.88968420 i$ \\
$10$ & 
$6.88156796 - 0.36773408 i$ & 
$6.88230126 - 0.36838601 i$ & 
$6.88449386 - 0.37033762 i$ & 
$6.88812433 - 0.37357663 i$ \\
$20$ & 
$12.75994862 - 0.13788816 i$ & 
$12.76068546 - 0.13825105 i$ & 
$12.76288803 - 0.13933865 i$ & 
$12.76653284 - 0.14114771 i$  \\
$30$ & 
$18.64042583 - 0.03916775 i$ & 
$18.64122199 - 0.03933376 i$ & 
$18.64360088 - 0.03983262 i$ & 
$18.64753421 - 0.04066680 i$  \\\hline
$\ell$ & $mM = 0.20 $ & $mM = 0.25 $ & $mM = 0.30 $ & $mM = 0.35 $ \\\hline
$0$ &
$1.48349180 - 1.16198028 i$ & 
$1.49377204 - 1.17351886 i$ & 
$1.50615298 - 1.18742468 i$ & 
$1.52053704 - 1.20359279 i$  \\
$1$ &
$1.83046511 - 1.03322029 i$ & 
$1.83943909 - 1.04417142 i$ & 
$1.85025488 - 1.05737742 i$ & 
$1.86283146 - 1.07274292 i$  \\
$2$ &
$2.31836207 - 0.89761108 i$ & 
$2.32638124 - 0.90769585 i$ & 
$2.33604694 - 0.91986473 i$ & 
$2.34728717 - 0.93403383 i$  \\
$10$ & 
$6.89315806 - 0.37808311 i$ & 
$6.89954897 - 0.38383036 i$ & 
$6.90724147 - 0.39078590 i$ & 
$6.91617272 - 0.39891264 i$  \\
$20$ & 
$12.77158207 - 0.14367296 i$ & 
$12.77798542 - 0.14690720 i$ & 
$12.78568235 - 0.15084152 i$ & 
$12.79460465 - 0.15546547 i$  \\
$30$ & 
$18.65297651 - 0.04184025 i$ & 
$18.65986742 - 0.04335825 i$ & 
$18.66813447 - 0.04522712 i$ & 
$18.67769632 - 0.04745393 i$  \\\hline
$\ell$ & $mM = 0.40 $ & $mM = 0.45 $ & $mM = 0.50 $ & $mM = 0.60 $ \\\hline
$0$ & 
$1.53681870 - 1.22190939 i$ & 
$1.55488782 - 1.24225525 i$ & 
$1.57463255 - 1.26450882 i$ & 
$1.61870707 - 1.31425634 i$  \\
$1$ & 
$1.87708111 - 1.09016436 i$ & 
$1.89291216 - 1.10953318 i$ & 
$1.91023135 - 1.13073850 i$ & 
$1.94896518 - 1.17821766 i$  \\
$2$ & 
$2.36002403 - 0.95011207 i$ & 
$2.37417610 - 0.96800382 i$ & 
$2.38966064 - 0.98761134 i$ & 
$2.42429982 - 1.03158437 i$  \\
$10$ & 
$6.92627490 - 0.40817005 i$ & 
$6.93747734 - 0.41851531 i$ & 
$6.94970842 - 0.42990431 i$ & 
$6.97697454 - 0.45563597 i$  \\
$20$ & 
$12.80467891 - 0.16076733 i$ & 
$12.81582892 - 0.16673431 i$ & 
$12.82797782 - 0.17335278 i$ & 
$12.85497129 - 0.18848669 i$  \\
$30$ & 
$18.68846579 - 0.05004619 i$ & 
$18.70035289 - 0.05301153 i$ & 
$18.71326738 - 0.05635742 i$ & 
$18.74182840 - 0.06421819 i$  \\\hline
\end{tabular}}
\end{table}

\begin{table}[ht]
\caption{Fundamental quasinormal frequencies $\omega M$ ($n=0$) of massive scalar fields with $\ell=0,1,2$ in the background of  Schwarzschild-AdS black holes with $M^2\Lambda =-0.005$.}
\label{T721}\centering
\scalebox{0.7}{
\begin{tabular}{ | c | c | c | c | c | c | c |c |}
\hline
$\ell$ & $mM = 0 $ & $mM = 0.05 $ & $mM = 0.10 $ & $mM = 0.15 $ \\\hline
$0$ & 
$0.11234408 - 0.00253387 i $ & 
$0.12751958 - 0.00347144 i$ & 
$0.15948982 - 0.00612361 i$ &  
$0.19628098 - 0.01039618 i$  \\
$1$ &
$0.15689207 - 0.00002564 i $ & 
$0.17300519 - 0.00004312 i$ &
$0.20700088 - 0.00011796 i$ &  
$0.24584844 - 0.00033499 i$  \\
$2$ & 
$0.19924295 - 7.07258*10^{-8} i$ & 
$0.21579949 - 1.3258*10^{-7} i$ &  
$0.25094984 - 4.4974*10^{-7} i$ & 
$0.29155767 - 1.6262*10^{-6} i$   \\\hline
%$3$ & 
%$0.316229907 -6.07595*10^{-9} i$ &
%$...$ & 
%$0.2932695192 - 1.11596*10^{-9} i$ & 
%$...$  \\
%$4$ & 
%$0.433022952 - 4.32018*10^{-10} i$ &  
%$...$ & 
%$0.4101012419 - 1.10042*10^{-10} i$ & 
%$...$   \\
%$5$ & 
%$0.622968180 - 5.67021*10^{-10} I$ &  
%$...$ & 
%$0.6003698070 - 1.9213*10^{-10} i$ & 
%$...$   \\\hline
$\ell$ & $mM = 0.20 $ & $mM = 0.25 $ & $mM = 0.30 $ & $mM = 0.35 $ \\\hline
$0$ &
$0.23466144 - 0.01622251 i$ & 
$0.27386451 - 0.02346464 i$ & 
$0.31372182 - 0.03195017 i$ &  
$0.35422112 - 0.04150833 i$   \\
$1$ &
$0.28556635 - 0.00088357 i$ &  
$0.32491192 - 0.00208114 i$ & 
$0.36356073 - 0.00429995 i$ & 
$0.40161853 - 0.00781653 i$  \\
$2$ &
$0.33367650 - 5.62773*10^{-6} i$ & 
$0.37605040 - 0.00001838 i$ & 
$0.41815083 - 0.00005661 i$ & 
$0.45967430 - 0.00016397 i$  \\\hline
%$3$ & 
%$0.2576012662 - 2.79494*10^{-10} i$ & 
%$...$ & 
%$0.464775312 -2.61267*10^{-7} I$ & 
%$...$   \\
%$4$ & 
%$0.3748430288 - 2.79796*10^{-11} i$ &  
%$...$ & 
%$0.4407514006 - 5.82734*10^{-10} i$ & 
%$...$  \\
%$5$ & 
%$0.5658747160 - 5.79203*10^{-11} i$ & 
%$...$ & 
%$0.6305500508 - 7.49728*10^{-10} i$ & 
%$...$  \\\hline
%$\ell$ & $mM = 0.40 $ & $mM = 0.45 $ & $mM = 0.50 $ & $mM = 0.60 $ %\\\hline
%$0$ & 
%$0.11654214403 - 0.00277397923 i$ &  
%$...$ & 
%$...$ & 
%$0.866089748 - 0.020264996 I$  \\
%$1$ & 
%$...$ & 
%$...$ & 
%$...$ &  
%$0.541462114 - 0.021000348 I$   \\
%$2$ & 
%$...$ & 
%$...$ &  
%$...$ & 
%$0.603870745 - 0.009262185 I$  \\\hline
%%$3$ & 
%$...$ & 
%$...$ & 
%$...$ &  
%$0.2429366656 - 1.50218*10^{-10} i$  \\
%$4$ & 
%$...$ & 
%$...$ & 
%$...$ & 
%$0.4350529863 - 4.6758*10^{-10} i$   \\
%$5$ & 
%$...$ & 
%$...$ & 
%$...$ & 
%$0.6249595063 - 6.10374*10^{-10} i$   \\\hline
\end{tabular}}
\end{table}

\clearpage

\acknowledgments
 We thank the referee for his/her careful review of the manuscript and his/her valuable comments and suggestions which helped us to improve the manuscript considerably. P. A. G. acknowledges the hospitality of the Universidad de La Serena where part of this work was undertaken. Y.V. acknowledge support by the Direcci\'on de Investigaci\'on y Desarrollo de la Universidad de La Serena, Grant No. PR18142.

\end{document}